\documentclass[aoas,MSNbibl,nameyear,seceqn,dvips]{arximspdf}
\usepackage{mathbh,dcolumn}
\usepackage{graphicx}
\doi{10.1214/12-AOAS587} 
\volume{6}
\issue{4}
\pubyear{2012}
\firstpage{1552}
\lastpage{1587}

\makeatletter
\newcolumntype{d}[1]{D{.}{.}{#1}}

\newcommand{\rright}{\right}
\newcommand{\lleft}{\left}
\newtheorem{theorem}{Theorem}[section]
\newproclaim{remark}[theorem]{Remark}
\def\eqref#1{(\ref{#1})}

\newcommand{\bX}{\bar{X}}
\newcommand{\bfe}{\mathbf{e}}
\newcommand{\bfp}{\mathbf{p}}

\newcommand{\bfw}{\mathbf{w}}
\newcommand{\bfX}{\mathbf{X}}
\newcommand{\bfY}{\mathbf{Y}}
\newcommand{\bfZ}{\mathbf{Z}}
\newcommand{\bfone}{\mathbf{1}}
\newcommand{\bbr}{\mathbb{R}}

\newcommand{\bbx}{\mathbb{X}}
\newcommand{\bbz}{\mathbb{Z}}
\newcommand{\F}{\mathcal{F}}
\newcommand{\R}{\mathcal{R}}
\newcommand{\X}{\mathcal{X}}

\newcommand{\al}{\alpha}
\newcommand{\be}{\beta}
\newcommand{\De}{\Delta}
\newcommand{\ep}{\varepsilon}
\newcommand{\Ga}{\Gamma}
\newcommand{\La}{\Lambda}
\newcommand{\Om}{\Omega}
\newcommand{\Si}{\Sigma}
\newcommand{\ta}{\tau}
\newcommand{\te}{\theta}

\newcommand{\hth}{\hat{\theta}}

\newcommand{\tth}{{\tilde{\theta}}}

\newcommand{\cov}{\operatorname{Cov}}

\newcommand{\var}{\operatorname{Var}}

\newcommand{\ind}{\mathbh{1}}

\newcommand{\raw}{\rightarrow}

\newcommand{\hSi}{\hat{\Sigma}}

\makeatother

\begin{document}
\begin{frontmatter}

\title{Gap bootstrap methods for massive data sets with an application
to transportation engineering\thanksref{T1}}
\thankstext{T1}{Supported in part by grant from
the U.S. Department of Transportation, University Transportation Centers
Program to the University Transportation Center for Mobility (DTRT06-G-0044)
and by NSF Grants no. DMS-07-07139 and DMS-10-07703.}
\runtitle{Gap bootstrap for massive data sets}

\begin{aug}
\author[A]{\fnms{S.~N.} \snm{Lahiri}\corref{}\ead[label=e1]{snlahiri@stat.tamu.edu}},
\author[A]{\fnms{C.} \snm{Spiegelman}},
\author[B]{\fnms{J.} \snm{Appiah}}
\and
\author[B]{\fnms{L.} \snm{Rilett}}
\runauthor{Lahiri, Spiegelman, Appiah and Rilett}
\affiliation{Texas A\&M University, Texas A\&M University,
University of Nebraska-Lincoln and University of Nebraska-Lincoln}
\address[A]{S.~N. Lahiri\\
C. Spiegelman\\
Department of Statistics\\
Texas A\&M University\\
College Station, Texas 77843\\
USA\\
\printead{e1}} 
\address[B]{J. Appiah\\
L. Rilett\\
Nebraska Transportation Center\\
University of Nebraska-Lincoln\\
262D Whittier Research Center\\
P.O. Box 880531\\
Lincoln, Nebraska 68588-0531\\
USA}
\end{aug}

\received{\smonth{4} \syear{2012}}

%
\begin{abstract}
In this paper we describe two bootstrap methods for massive data sets.
Naive applications of common resampling methodology are often
impractical for massive data sets due to
computational burden and due to complex
patterns of inhomogeneity. In contrast, the
proposed methods exploit certain structural properties of a large class
of massive data sets to break up the original problem into a set of
simpler subproblems, solve each subproblem separately where the
data exhibit approximate uniformity and where
computational complexity can be reduced 
to a manageable level, and then combine the
results through certain analytical considerations.
The validity of the proposed methods is proved and their finite sample
properties are studied through a moderately large
simulation study. The methodology is
illustrated with a real data example from
Transportation Engineering, which motivated the development of the
proposed methods.
\end{abstract}

%

\end{frontmatter}

\section{Introduction}\label{sec1}
Statistical analysis and inference for massive data sets present
unique challenges. Naive applications of standard statistical
methodology often become impractical, especially due to
increase in computational complexity.
While large data size is desirable from a statistical
inference perspective,
suitable modification of existing statistical methodology is needed
to handle such challenges associated with massive data sets.
In this paper, we propose a novel resampling methodology,
called the \textit{Gap Bootstrap},
for a large class of massive data sets
that possess certain structural properties. The proposed methodology
cleverly exploits the data structure
to break up the original inference problem
into smaller parts, use standard resampling methodology to each part
to reduce the computational complexity, and then use some analytical
considerations to put the individual pieces
together, thereby alleviating
the computational issues associated with large data sets to a great
extent.

The class of problems we consider here is the estimation of
standard errors of estimators of population parameters
based on massive multivariate data sets that may have
heterogeneous distributions. A primary example is the
origin-destination (OD) model in transportation
engineering. In an OD model, which motivates this work and which is
described in detail in Section~\ref{sec2} below, the data
represent traffic volumes at a number of origins and
destinations collected over short intervals of time
(e.g., 5 minute intervals) daily, over a long period
(several months), thereby leading to a massive data set.
Here, the main
goals of statistical analysis are (i) uncertainty
quantification associated
with the estimation of the parameters in the OD model
and (ii) to improve prediction of traffic volumes at
the origins and the destinations over
a given stretch of the highway.
Other examples of massive data sets having the
required structural property
include (i)
receptor modeling in environmental monitoring,
where spatio-temporal data are
collected for many pollution receptors over
a long time, and (ii)
toxicological models for dietary intakes and drugs,
where blood levels of a large number of
toxins and organic compounds
are monitored in repeated samples for a large number of patients.
The key feature of these data sets is the presence of ``gaps''
which allow one to partition the original data set
into smaller subsets with nice properties.

The ``largeness'' and potential inhomogeneity of such data sets
present challenges for estimated
model uncertainty evaluation. The standard propagation of
error formula or the delta method relies on assumptions of
independence and identical distributions,
stationarity (for space--time data)
or other kinds of uniformity which, in most instances,
are not appropriate for such data sets.
Alternatively, one may try to apply the
bootstrap and other resampling methods to assess the uncertainty.
It is known that the ordinary bootstrap method typically underestimates
the standard error for parameters when the data are dependent
(positively correlated).
The block bootstrap has become a popular tool for dealing with
dependent data. By using blocks, the
local dependence structure in the data is maintained and, hence, the
resulting estimates from the block bootstrap tend to be less
biased than those from the traditional (i.i.d.) bootstrap.
For more details, see Lahiri (\citeyear{Lah99,Lah03}).
However, computational complexity of naive block bootstrap
methods increases significantly with the size of the data sets, as the
given estimator has to be computed \textit{repeatedly} based on resamples
that have the same size as the original data set.
In this paper, we propose two resampling
methods, generally both referred to as
\textit{Gap Bootstraps}, that exploit the ``gap'' in the dependence
structure of such large-scale data sets to reduce the computational
burden.
%
Specifically, the gap bootstrap estimator of the standard error
is appropriate for data that can be
partitioned into approximately exchangeable or
homogeneous 
subsets. While
the distribution of the entire data set
is not exchangeable or homogeneous,
it
is entirely reasonable that many multivariate subsets will be
exchangeable or homogeneous. 
If the estimation method that is
being used is accurate,
then we show that the gap bootstrap gives a consistent and
asymptotically unbiased estimate of standard errors.
The \textit{key idea is to employ the bootstrap method to each
of the homogeneous 
subsets of the data
separately and then combine
the estimators from different subsets in a suitable way
to produce a valid estimator of the standard error of a
given estimator based on the entire
data set}. The proposed method is computationally much simpler
than the existing resampling methods that require
repeated computation of the original estimator,
which may not be feasible simply due to computational complexity
of the original estimator, at the scale of the whole data set.

The rest of the paper is organized as follows. In Section~\ref{sec2}
we describe the OD model and the data structure that motivate the
proposed methodology. In Section~\ref{sec3} we give the descriptions of
two variants of the Gap Bootstrap. Section~\ref{sec4} asserts
consistency of the
proposed Gap Bootstrap variance estimators. In Section~\ref{sec5}
we report results from a moderately large simulation study, which shows
that the proposed methods attain high levels of accuracy
for moderately large data sets under
various types of gap-dependence structures. In Section~\ref{sec6}
we revisit the OD models and apply the methodology to a real
data set from a study of traffic patterns, conducted by an
intelligent traffic management system on a test bed in San Antonio, TX.
Some concluding remarks are made in Section~\ref{sec7}.
Conditions for the validity
of the theoretical results
and outlines of the proofs
are given in
the \hyperref[app]{Appendix}.

\section{The OD models and the estimation problem}\label{sec2}
\subsection{Background}\label{sec2.1}
The key component of an origin-destination (OD) mod\-el
is an OD trip matrix that
reflects
the volume of traffic (number of trips, amount of freight, etc.)
between all possible origins and destinations in a transportation
network over a given time interval. The OD matrix can be measured directly,
albeit with much effort and at great costs, by conducting
individual interviews, license plate surveys, or by taking
aerial photographs [cf. Cramer and Keller (\citeyear{CreKel87})].
Because
of the cost involved in collecting direct measurements to populate a
traffic matrix, there has been considerable effort in recent years to
develop synthetic techniques which provide ``reasonable'' values for the
unknown OD matrix entries in a more indirect way, such as
using observed data from link volume counts from
inductive loop detectors. Over the past two
decades, numerous approaches to synthetic OD matrix estimation have
been proposed [Cascetta (\citeyear{Cas84}), Bell (\citeyear{Bel91}),
Okutani (\citeyear{Oku87}),
Dixon and Rilett (\citeyear{DixRil00})]. One common approach for estimating
the OD matrix from link volume counts is based on the
least squares regression where the unknown OD matrix
is estimated by minimizing the squared Euclidean distance
between the observed link and the estimated link
volumes.

\subsection{Data structure}\label{sec2.2}
\begin{figure}

\includegraphics{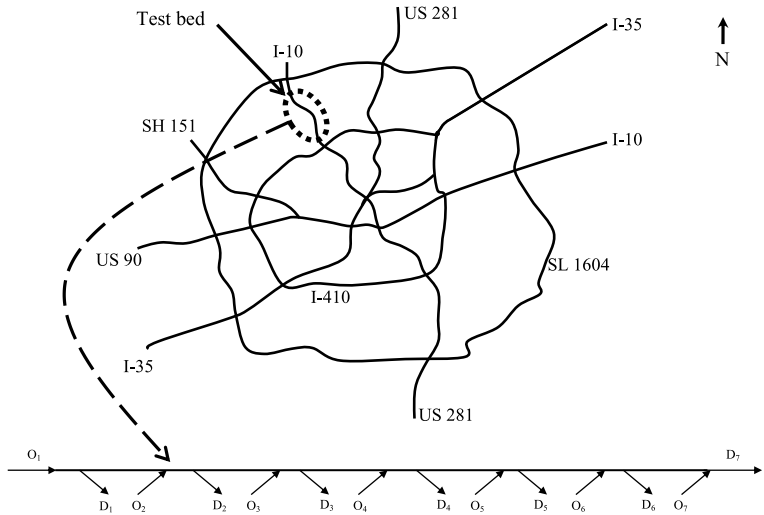}

\caption{The transportation network in San Antonio, TX under
study.}\label{fig1}\vspace*{-3pt}
\end{figure}

The data are in the form of a time series of link volume counts
measured at several on/off ramp locations on a freeway using an
inductive loop detector, such as in Figure~\ref{fig1}.

%

Here $O_k$ and $D_k$, respectively,
represent the traffic volumes at the $k$th origin
and the $k$th destination over a given stretch of a highway.
The analysis period is divided into $T$ time
periods of equal duration $\Delta t$. The time series of
link volume counts is generally \textit{periodic}
and \textit{weakly
dependent}, that is, the dependence dies off as the separation
of the time intervals becomes large. For example, daily data over
each given time slot of duration $\Delta t$ are similar,
but data over well separated time slots (e.g., time slots in
Monday morning and Monday afternoon) can be different.
This implies that the traffic data have a periodic structure.
Further, Monday at 8:00--8:05 am
data have nontrivial correlation with Monday at 8:05--8:10 am data,
but neither data set says anything about Tuesday data
at 8:00--8:05 am (showing \textit{approximate independence}).
Accordingly, let $\bfY_t, t=1,2\ldots,$ be a $d$-dimensional
time series, representing the link volume counts
at a given set of on/off ramp locations over the $t$th
time interval. Suppose that we are interested in reconstructing
the OD matrix for $p$-many
short intervals during the morning rush hours, such as
$36$ link volume counts over $\Delta t = 5$-minute intervals,
extending from 8:00 am through 11:00 am,\vadjust{\goodbreak} at several on/off
ramp locations.
Thus, the observed data for the OD modeling is a part of the
$\bfY_t$ series,
\[
\{\bfX_1,\ldots, \bfX_p; \ldots; \bfX_{(m-1)p+1},
\ldots, \bfX_{mp}\},
\]
where the link volume counts are
observed over the $p$-intervals on each day,
for~$m$ days, giving a $d$-dimensional multivariate sample
of size $n= mp$. There are $q=T-p$ time slots between the
last observation on any given day and the first observation
on the next day, which introduces the ``gap'' structure in
the $\bfX_t$-series. Specifically, in terms of the $\bfY_t$-series, the
$\bfX_t$-variables are given by
\[
\bfX_{ip+j} = \bfY_{i(p+q)+j},\qquad j=1,\ldots,p, i=0,\ldots, m-1.
\]
For data collected over a large
transportation network and over a long period of time, $d$ and
$m$ are large, leading to a massive data set.
Observe that the $\bfX_t$-variables can be arranged in a $p\times m$
matrix, where each element of the matrix-array gives a $d$-dimensional
data value:
%
%
\begin{equation}\label{data-array}
\bbx= \pmatrix{ \bfX_1 &
\bfX_{p+1} &\ldots& \bfX_{(m-1)p+1}
\vspace*{2pt}\cr
\bfX_2 & \bfX_{p+2} &\ldots& \bfX_{(m-1)p+2}
\vspace*{2pt}\cr
\cdot&\cdot&\ldots&\cdot
\vspace*{2pt}\cr
\cdot&\cdot&\ldots&\cdot
\vspace*{2pt}\cr
\bfX_{p} & \bfX_{2p} &\ldots& \bfX_{mp}}.
\end{equation}
Due to the arrangement of the $p$ time slots in the $j$th day
along the $j$th column in \eqref{data-array}, the rows
in the array \eqref{data-array} correspond
to a fixed time slot over days and are expected to exhibit
a similar distribution of the link volume counts;
although a day-of-week variation might be present, the
standard practice in the Transportation engineering
is to treat the weekdays as similar [cf. Roess, Prassas and McShane
(\citeyear{RoePraMcS04}), Mannering, Washburn and
Kilareski (\citeyear{ManWasKil09})].
On the other hand,
due to the ``gap'' between the last time slot on
the $j$th day and the first time slot of the $(j+1)$st day,
the variables in the $j$th and the $(j+1)$st columns are
essentially independent. Hence, this yields a data structure
where
%
%
\begin{equation}
\lleft. %
\begin{array} {ll} \textup{(a)}& \mbox{the variables within each
column have serial correlations}
\\
& \mbox{and possibly nonstationary distributions,}
\\
\textup{(b)} & \mbox{the variables in each row are identically distributed, and}
\\
\textup{(c)}& \mbox{the columns are approximately independent arrays}\\
&\mbox{of random vectors.}
\end{array}
\rright\} \label{proc-c}
\end{equation}

In the transportation engineering application, each random vector $\bfX
_t$ represents the
link volume counts in
a transportation network corresponding to
$r$ origin (entrance) ramps and $s$ destination (exit) ramps
as shown in Figure~\ref{fig1}.
%
%
Let $o_{\ell t}$ and $d_{kt}$, respectively, denote the
link volumes at origin $\ell$ and at
destination $k$ at time~$t$. Then the components of $\bfX_t$
for each $t$ are given by the
$d\equiv(r+s)$-variables\vadjust{\goodbreak}
$\{o_{\ell t}\dvtx\ell=1,\ldots, r\}\cup\{d_{kt}\dvtx k=1,\ldots, s\}$.
Given the link volume counts on all origin and destination ramps, the
fraction $p_{k\ell}$ (known as the
\textit{OD split proportion}) of vehicles that exit the
system at destination ramp $k$
given that they entered at origin ramp $\ell$
can be calculated. This
is because the link volume at
destination $k$ at time~$t$, $d_{kt}$,
is a linear combination of the OD split
proportions and the origin volumes at time $t$, $o_{\ell t}$'s.
In the synthetic OD model, $p_{k\ell}$'s
are the \textit{unknown} system parameters and have to be estimated.
Once the split proportions are available, the OD matrix for each
time period can be identified as a linear combination of the
split proportion matrix and the vector of origin volumes.
The key statistical inference issue here is to quantify the size of the
standard errors of the estimated split proportions
in the synthetic OD model.\

\section{Resampling methodology}\label{sec3}
\subsection{Basic framework}\label{sec3.1}
\setcounter{equation}{0}
To describe the resampling methodology, we adopt a framework
that mimics the ``gap structure'' of the OD model in Section~\ref{sec2}.
Let
$\{\bfX_1,\ldots, \bfX_p; \ldots; \bfX_{(m-1)p+1},\ldots, \bfX_{mp}\}$
be a $d$-dimensional time series with
stationary components $\{\bfX_{ip+j}\dvtx i=0,\ldots, m-1\}$ for
$j=1,\ldots, p$ such that the corresponding array
\eqref{data-array} satisfies \eqref{proc-c}. For example, such
a time series results from a periodic, multivariate parent
time series $\bfY_t$ that is $m_0$-dependent for some $m_0\geq0$
and that is observed with ``gaps'' of length $q>m_0$.
In general, the dependence structure of the original time series
$\bfY_t$ is retained within each complete period
$\{\bfX_{ip+j}\dvtx j=1,\ldots, p\}$, $i=0,\ldots,m$, but the random variables
belonging to two different periods are essentially independent. Let
$\te$
be a vector-valued parameter of interest and let ${\hat{\theta}}_{n}$
be an estimator of $\te$ based on $\bfX_1,\ldots\bfX_n$,
where $n=mp$ denotes the sample size. We
now formulate two resampling methods for estimating the standard error
of ${\hat{\theta}}_{n}$ that are suitable for massive data sets with such
``gap'' structures. The first method is applicable when the $p$ rows of the
array \eqref{data-array} are \textit{exchangeable} and the second one
is applicable where the rows are possibly nonidentically distributed
and
where the variables within each column
have serial dependence.

\subsection{Gap Bootstrap I}\label{sec3.2}
Let $\bfX_{(j)} = (\bfX_{ip+j}\dvtx i=0,\ldots, m-1)$
denote the $j$th row of the array $\bbx$ 
in
\eqref{data-array}.
For the time being, assume that the rows of $\bbx$ are
exchangeable, that is, for any permutation $(j_1,\ldots,j_p)$
of the integers
$(1,\ldots,p)$,
$\{\bfX_{(j_1)},\ldots, \bfX_{(j_p)}\}$
have the same joint distribution as
$\{\bfX_{(1)},\ldots,\break \bfX_{(p)}\}$, although we do not need the full
force of exchangeability for the validity of the method (cf. Section
\ref{sec4}).
For notational compactness, set $\bfX_{(0)}=\bbx$.
Next suppose that the parameter $\te$ can be estimated by
using the row variables $\bfX_{(j)}$ as well as using the
complete data set, through estimating equations of the
form
\[
\Psi_j(\bfX_{(j)}; \te) =0,\qquad j=0,1,\ldots, p,\vadjust{\goodbreak}
\]
resulting in the estimators $\hth_{jn}$,
based on the $j$th row, for $j=1,\ldots,p$, and
the
estimator ${\hat{\theta}}_{n}=\hth_{0n}$ for $j=0$ based on
the entire data set, respectively. It is obvious that for
large values of $p$, the computation of $\hth_{jn}$'s
can be much simpler than that of ${\hat{\theta}}_{n}$, as the estimators
$\hth_{jn}$'s are based on a fraction
(namely, $\frac{1}{p}$)
of the
total observations.
On the other hand, the individual $\hth_{jn}$'s
lose efficiency, as they are based on a subset of the data.
However, under some mild conditions
on the score functions, the M-estimators
can be asymptotically linearized by using the
averages of the influence functions over the
respective data sets $\bfX_{(j)}$
[cf. Chapter 7, Serfling (\citeyear{Ser80})]. As a result,
under such regularity conditions,
%
%
\begin{equation}
\bar{\te}_n \equiv p^{-1}\sum
_{j=1}^p \hth_{jn} \label{th-ave}
\end{equation}
gives an asymptotically equivalent
approximation to ${\hat{\theta}}_{n}$. Now an estimator of
the variance of the original estimator ${\hat{\theta}}_{n}$ can be
obtained by combining the variance estimators of the
$\hth_{jn}$'s through the equation
%
%
\begin{equation}
\var(\bar{\te}_n) = p^{-2} \Biggl[\sum
_{j=1}^p \var(\hth_{jn}) + \sum
_{1\leq j\neq k \leq p} \cov( \hth_{jn}, \hth_{kn}) \Biggr].
\label{lin-app}
\end{equation}
Note that using the i.i.d. assumption on the row variables,
an estimator of $\var(\hth_{jn})$ can be found by the ordinary
bootstrap method (also referred to as the \textit{i.i.d. bootstrap} in here)
of Efron (\citeyear{Efr79})
that selects a with replacement sample
of size $m$ from the $j$th row of data values.
We denote this by
$\widehat{\var}(\hth_{jn})$ (and also by $\hSi_{jn}$), $j=1,\ldots
, p$.
Further, under the exchangeability
assumption, all the covariance terms are equal and, hence,
we may estimate the cross-covariance terms by
estimating the variance of the pairwise
differences as follows:
\[
\widetilde{\var}(\hth_{j_0n}-\hth_{k_0n}) = \frac{\sum_{1\leq j\neq k
\leq p} (\hth_{jn}- \hth_{kn})
(\hth_{jn}- \hth_{kn})'}{p(p-1)}, \qquad 1
\leq j_0\neq k_0 \leq p.
\]
%
Then, the cross covariance estimator is given by
\[
\widetilde{\cov}(\hth_{j_0n},\hth_{k_0n}) = \bigl[
\hSi_{j_0n}+ \hSi_{k_0n} -\widetilde{\var}(\hth_{j_0n}-
\hth_{k_0n}) \bigr]/2.
\]
Plugging in these estimators of the variance and the covariance terms
in \eqref{lin-app}
yields the \textit{Gap Bootstrap Method I estimator of the variance} of
${\hat{\theta}}_{n}$
as
%
%
\begin{equation}\label{gp-I}
\widehat{\var}_{\mathrm{GB}\mbox{-}\mathrm{I}}({\hat{\theta}}_{n}) = p^{-2}
\Biggl[\sum_{j=1}^p \widehat{\var}(
\hth_{jn}) + \sum_{1\leq j\neq k \leq p} \widetilde{\cov}(
\hth_{jn}, \hth_{kn}) \Biggr].
\end{equation}

Note that the estimator proposed here only requires computation of
the parameter
estimators based on the $p$ subsets, which can cut down on the computational
complexity significantly when $p$ is large.\vadjust{\goodbreak}

\subsection{Gap Bootstrap II}\label{sec3.3}
In this section we describe a Gap Bootstrap method for the more general
case where the rows $\bfX_{(j)}$'s in \eqref{data-array} are not necessarily
exchangeable and, hence, do not have the same distribution.
Further, we allow the columns of $\bbx$ to
have certain serial dependence.
This, for example, is the situation when
the $\bfX_t$-series is obtained from a weakly dependent parent series
$\{\bfY_t\}$ by systematic deletion of
$q$-components, creating the ``gap'' structure in the
observed $\bfX_t$-series as described in Section~\ref{sec2}.
If the $\bfY_t$-series is $m_0$-dependent
with an $m_0< q$,
then $\{\bfX_t\}$ satisfies the conditions in \eqref{proc-c}. For a
mixing sequence $\bfY_t$, the gapped segments are never exactly
independent, but the effect of the dependence on
the gapped segments are practically negligible for large enough
``gaps,'' so that approximate independence of the columns
holds when $q$ is large.
We restrict attention to the simplified structure \eqref{proc-c} to
motivate the main ideas and to keep
the exposition simple. Validity of the theoretical
results continue to hold under weak dependence
among the columns of the array \eqref{data-array};
see Section~\ref{sec4} for further details.

As in the case of Gap Bootstrap I, we suppose that
the parameter $\te$ can be estimated by
using the row variables $\bfX_{(j)}$ as well as using the
complete data set, resulting in the estimator
$\hth_{jn}$,
based on the $j$th row for $j=1,\ldots,p$ and
the estimator ${\hat{\theta}}_{n}=\hth_{0n}$ (for $j=0$) based on
the entire data set, respectively. The estimation method
can be any standard method, including those based on
score functions and quasi-maximum likelihood methods, such that the following
\textit{asymptotic linearity condition}
holds:

\textit{There exist known weights $w_{1n},\ldots,w_{pn}\in[0,1]$ with
$\sum_{j=1}^pw_{jn}=1$ such that }
%
%
\begin{equation}
{\hat{\theta}}_{n}- \sum_{j=1}^p
w_{jn}\hth_{jn} = o_P\bigl(n^{-1/2}
\bigr)\qquad\mbox{as }n\rightarrow\infty. \label{eqn-lin-rep}
\end{equation}

Classes of such estimators are given by (i) L-, M- and R-estimators
of location parameters [cf. Koul and Mukherjee (\citeyear{KouMuk93})], (ii)
differentiable functionals of the (weighted) empirical process
[cf. Serfling (\citeyear{Ser80}), Koul (\citeyear{Kou02})],
and (iii) estimators satisfying the smooth function model
[cf. Hall (\citeyear{ha92}), Lahiri (\citeyear{Lah03})].
%
An explicit example of an estimator satisfying
\eqref{eqn-lin-rep} is given in Remark 3.5 below
[cf. \eqref{row-rep}] and the details of verification
of \eqref{eqn-lin-rep} are given in the \hyperref[app]{Appendix}.

Note that under (\ref{eqn-lin-rep}), the asymptotic
variance of $n^{1/2}({\hat{\theta}}_{n}-\te)$ is given by the asymptotic
variance of $\sum_{j=1}^p w_{jn}n^{1/2}(\hth_{jn} -\te)$. The latter
involves both variances and covariances of the row-wise
estimators $\hth_{jn}$'s.
%
The Gap Bootstrap method II estimator of the variance
of ${\hat{\theta}}_{n}$ is obtained by combining
individual variance estimators of the marginal estimators $\hth_{jn}$'s
with estimators of their cross covariances.
Note that as the row-wise estimators $\hth_{jn}$
are based on (approximately) i.i.d. data, as in the case of
Gap Bootstrap method I,
one can use the i.i.d. bootstrap method of Efron (\citeyear{Efr79})
\textit{within} each row $\bfX_{(j)}$ and obtain an
estimator of the standard error of
each $\hth_{jn}$. We continue to denote these by $\widehat{\var
}(\hth_{jn})$,
$1\leq j\leq p$, as in Section~\ref{sec3.2}. However, since we now allow
the presence of temporal
dependence \textit{among} the rows, resampling individual observations
is not enough [cf. Singh (\citeyear{Sin81})] for cross-covariance estimation
and some version of block resampling
is needed [cf. K\"unsch (\citeyear{Kun89}), Lahiri (\citeyear{Lah03})].
As explained earlier, repeated computation of the estimator ${\hat
{\theta}}_{n}$
based on replicates of the full sample may not be feasible
merely due to the associated computational costs.
Instead, computation of the replicates on
smaller portions of the data may be much faster
(as it avoids \textit{repeated} resampling)
and stable. This
motivates us to
consider
the sampling window method of Politis and Romano (\citeyear{PolRom94}) and
Hall and Jing (\citeyear{HalJin96}) for
cross-covariance estimation. Compared to the block bootstrap
methods, the sampling window method is computationally much faster
but at the same time, it typically achieves
the same level of accuracy as the block bootstrap covariance estimators,
asymptotically [cf. Lahiri (\citeyear{Lah03})].
The main steps of the Gap Bootstrap Method II are as follows.

\subsubsection{The univariate parameter case}\label{sec3.3.1}
For simplicity, we first describe the steps of the Gap Bootstrap Method II
for the case where the parameter $\te$ is
\textit{one-dimensional}:

\textit{Steps}:
\begin{enumerate}[(III)]
\item[(I)] Use i.i.d. resampling of individual observations within each row
to construct a bootstrap estimator $\widehat{\var}(\hth_{jn})$
of $\var(\hth_{jn})$, $j=1,\ldots,p,$ as in the case of
Gap Bootstrap method I.
In the one-dimensional case,
we will denote these by $\hat{\sigma}^2_{jn}$, $j=1,\ldots,p$.

\item[(II)]
The Gap Bootstrap II estimator of the asymptotic
variance of ${\hat{\theta}}_{n}$ is given by
%
%
\begin{equation}
{\bar\ta}_n^2 = \sum_{j=1}^p
\sum_{k=1}^p w_{jn}w_{kn}
\hat{\sigma}_{jn}\hat{\sigma}_{kn} {\tilde
\rho}_{n}(j,k),
\end{equation}
where $\hat{\sigma}_{jn}^2$ is as in Step I and where
${\tilde{\rho}}_{n}(j,k)$ is the sampling window estimator of the
asymptotic correlation between
$\hth_{jn}$
and
$\hth_{kn}$, described below.

\item[(III)]
To estimate the correlation ${\rho}_{n}(j,k)$
between $\hth_{jn}$ and $\hth_{kn}$ by
the sampling window method
[cf. Politis and Romano (\citeyear{PolRom94}) and
Hall and Jing (\citeyear{HalJin96})], 
first fix
an integer $\ell\in(1,m)$.
Also, let
\begin{eqnarray*}
\bfX^{(1)} &=& (\bfX_{1},\ldots,\bfX_{p}),\qquad
\bfX^{(2)}=(\bfX_{p+1},\ldots,\bfX_{2p}),\ldots,\\
\bfX^{(m)}&=&(\bfX_{(m-1)p+1},\ldots,X_{mp})
\end{eqnarray*}
denote the columns of the matrix array \eqref{data-array}.
%
The version of the sampling window
method that we will employ here will be based on
(overlapping) subseries of $\ell$ columns.
%
%
The following are the
main steps of the sampling window method:

\begin{enumerate}[(IIIb)]
\item[(IIIa)]
Define the overlapping subseries of the column-variables $\bfX^{(\cdot)}$
of length $\ell$ as
\[
\X_i=\bigl(\bfX^{(i)},\ldots,\bfX^{(i+\ell-1)}\bigr),\qquad
i=1,\ldots,I,
\]
%
where $I=m-\ell+1$.
Note that each subseries $\X_i$ contains $\ell$ complete columns or periods
and consists of $\ell p$-many $\bfX_t$-variables.

\item[(IIIb)]
Next, for each $i=1,\ldots, I$,
we employ the given estimation algorithm to the
$\bfX_t$-variables in $\X_i$ to construct
the subseries version $\tth^{(i)}_{jn}$
of $\hth_{jn}$, $j=1,\ldots,p$.
(There is a
slight abuse of notation here, as the sample size for the
$i$th subseries of $\bfX_t$-variables is $\ell p$,
not $n=mp$ and, hence, we should be using $\tth^{(i)}_{j(\ell p)}$
instead of\vspace*{1pt}
$\tth^{(i)}_{jn}$, but we drop the more elaborate notation for simplicity).

\item[(IIIc)]
For
$1\leq j<k\leq p$,
the sampling window estimator of the correlation between
$\hth_{jn}$ and $\hth_{kn}$
is given by
%
%
\begin{equation}
{\tilde{\rho}}_{n}(j,k) = \frac{I^{-1}\sum_{i=1}^I (\tth_{jn}^{(i)}
-{\hat{\theta}}_{n}) %
(\tth_{kn}^{(i)} -{\hat{\theta}}_{n})} 
{
[I^{-1}\sum_{i=1}^I (
\tth_{jn}^{(i)} -{\hat{\theta}}_{n}
)^2 
]^{1/2} [I^{-1}
\sum_{i=1}^I (\tth_{kn}^{(i)}
-{\hat{\theta}}_{n})^2 
]^{1/2}}.
\end{equation}

\end{enumerate}
\end{enumerate}


\subsubsection{The multivariate parameter case}\label{sec3.3.2}
The multivariate version of the Gap bootstrap estimator of the variance matrix
of a vector parameter estimator ${\hat{\theta}}_{n}$ can be derived
using the same arguments, with routine
changes in the notation. Let $\hSi_{jn}$ denote the bootstrap estimator
of $\var({\hat{\theta}}_{jn})$, based on the i.i.d. bootstrap method of
Efron (\citeyear{Efr79}).
Next, with the subsampling replicates
$\tth^{(i)}_{jn}$, $j=1,\ldots,p$, based on the overlapping blocks
$\{\X_i\dvtx i=1,\ldots, I\}$ of $\ell$ columns each
(cf. Step [III] of Section~\ref{sec3.3.1}), define the sampling window
estimator
${\tilde{\R}}_{n}(j,k)$ of the correlation
matrix of $\hth_{jn}$ and $\hth_{kn}$ as
\begin{eqnarray*}
{\tilde{\R}}_{n}(j,k) &=& \Biggl[I^{-1}\sum
_{i=1}^I \bigl(\tth_{jn}^{(i)} -{
\hat{\theta}}_{n}\bigr) 
\bigl(\tth_{jn}^{(i)}
- {\hat{\theta}}_{n}\bigr)' 
\Biggr]^{-1/2}
\\
&&{} \times\Biggl\{I^{-1}\sum_{i=1}^I
\bigl(\tth_{jn}^{(i)} - {\hat{\theta}}_{n}\bigr)
\bigl(\tth_{km}^{(i)} - {\hat{
\theta}}_{n}\bigr)' 
\Biggr\}
\\
&&{}\times\Biggl[I^{-1}\sum_{i=1}^I
\bigl(\tth_{km}^{(i)} - {\hat{\theta}}_{n}\bigr)
\bigl(\tth_{km}^{(i)} -{\hat{\theta}}_{n}
\bigr)' 
\Biggr]^{-1/2}.
\end{eqnarray*}
%
%
Then the variance estimator
based on Gap bootstrap II is given by
%
%
\begin{equation}
\widehat{\var}_{\mathrm{GB}\mbox{-}\mathrm{II}}({\hat{\theta}}_{n}) =
\sum
_{j=1}^p \sum_{k=1}^p
w_{jn}w_{kn} \hSi_{jn}^{1/2}{\tilde{
\R}}_{n}(j,k)\hSi_{kn}^{1/2}. \label{gp-II-m}
\end{equation}

\subsubsection{Some comments on Method II}\label{sec3.3.3}

\begin{remark}\label{rem3.1} Note that for estimators $\{\tth_{jn}\dvtx
j=1,\ldots
,p\}$
with large asymptotic variances, estimation
of the correlation coefficients by the sampling window method
is more stable, as these are \textit{bounded} (and have a compact
support). On the other hand, the asymptotic variances of
$\hth_{jn}$'s have an unbounded range of values and therefore
are more difficult to estimate accurately. Since variance estimation by
Efron (\citeyear{Efr79})'s bootstrap has a higher level of accuracy
[e.g., $O_P(n^{-1/2})$] compared to the
sampling window method variance estimation [with
the slower rate $O_P([\ell/n]^{1/2} + \ell^{-1})$;
see Lahiri (\citeyear{Lah03})], the proposed approach is expected to
lead to a
better overall performance than a direct application of the
sampling window method
to estimate the variance of ${\hat{\theta}}_{n}$.
\end{remark}

\begin{remark}\label{rem3.2} Note that all estimators computed here
(apart from a one-time
computation of ${\hat{\theta}}_{n}$ in the sampling window method)
are based on subsamples
and hence are computationally simpler than repeated computation of
${\hat{\theta}}_{n}$
required by naive applications of the block resampling methods.
\end{remark}

\begin{remark}\label{rem3.3}
For applying Gap Bootstrap II, the user needs to specify the block
length $l$. Several standard block length selection
rules are available in the block resampling literature
[cf. Chapter 7, Lahiri (\citeyear{Lah03})] for estimating the
variance--covariance parameters. Any of these
are applicable in our problem. Specifically,
we mention the plug-in method of Patton, Politis and  White (\citeyear{PatPolWhi09}) that is computationally simple
and, hence, is specially suited for large data sets.
\end{remark}

\begin{remark}\label{rem3.4} The proposed estimator remains valid (i.e., consistent)
under more
general conditions
than \eqref{proc-c}, where the columns of the array \eqref
{data-array} are
not necessarily independent.
In particular, the proposed estimator in \eqref{gp-II-m}
remains consistent even when the $\bfX_t$
variables in the array \eqref{data-array} are obtained by
creating ``gaps'' in a
weakly dependent (e.g., strongly mixing) parent
time series $\bfY_t$.
This is because the subsampling window
method employed in the construction of the cross-correlation
can effectively capture the residual dependence
structure among the columns of the array \eqref{data-array}.
The use of i.i.d. bootstrap
to construct the variance estimators $\hSi_{jn}$ is adequate
when the gap is large, as
the separation of two consecutive random variables within
a row makes the correlation
negligible.
See Theorem~\ref{th4.2} below and its proof in the \hyperref[app]{Appendix}.
\end{remark}

\begin{remark}\label{rem3.5} An alternative, intuitive approach to estimating
the variance of ${\hat{\theta}}_{n}$ 
is to consider the data array
\eqref{data-array} by columns rather than by rows.
Let $\hth^{(1)},\ldots,\hth^{(m)}$ denote the estimates of $\te$
based on the $m$ \textit{columns} of the data matrix $\bbx$.
Then, assuming that the columns of $\bbx$ are (approximately)
independent and assuming that $\hth^{(1)},\ldots,\hth^{(m)}$
are identically distributed, one may be tempted to
estimate $\var({\hat{\theta}}_{n})$ by using the sample variance of the
$\hth^{(1)},\ldots,\hth^{(m)}$, based on the following analog of
\eqref{th-ave}:
%
%
\begin{equation}
{\hat{\theta}}_{n}\approx m^{-1}\sum
_{k=1}^m \hth^{(k)}. \label{naive-rep}
\end{equation}
\end{remark}
However, when $p$ is small compared to $m$, such an approximation
is sub-optimal, and this approach may
drastically fail
if $p$ is fixed. As an illustrating example, consider
the case where the $\bfX_i$'s are 1-dimensional random variables,
$p\geq1$ is fixed (i.e., it does not depend on the sample size), $n=mp$,
and the columns $\bfX^{(k)}, k=1,\ldots,m,$ have an
``identical distribution'' with mean vector $(\mu,\ldots,\mu)'\in
\bbr^p$
and $p\times p$ covariance matrix $\Si$.
For simplicity,
also suppose that the diagonal elements of $\Si$ are all equal to
$\sigma^2\in(0,\infty)$.
Let
\[
{\hat{\theta}}_{n}= n^{-1}\sum
_{i=1}^n (X_i-{\bar{X}}_n)^2,
\]
an estimator of $\te=p^{-1}$ trace$(\Si)=\sigma^2$.
Let $\hth^{(k)}$ and $\hth_{jn}$, respectively, denote the
sample variance of the $X_t$'s in the $k$th column and the
$j$th row, $k=1,\ldots,m$ and $j=1,\ldots,p$. Then, in Appendix~\ref{pf-t4.2}, we show that
%
%
\begin{equation}\label{row-rep}
{\hat{\theta}}_{n}= p^{-1}\sum
_{j=1}^p \hth_{jn} + o_p
\bigl(n^{-1/2}\bigr),
\end{equation}
while
\begin{equation}\label{col-rep}
{\hat{\theta}}_{n}= m^{-1}\sum
_{k=1}^m \hth^{(k)} + p^{-2}
\bfone'\Si\bfone+O_p\bigl(n^{-1/2}\bigr),
\end{equation}
where $\bfone$ is the $p\times1$ vector of $1$'s.
Thus, in this example, \eqref{eqn-lin-rep} holds with $w_{jn}=p^{-1}$
for $1\leq j\leq p$. However, \eqref{col-rep} shows that the column-wise
approach based on \eqref{naive-rep} results in a very
crude approximation which \textit{fails} to satisfy an analog
of~\eqref{eqn-lin-rep}.
For estimating the variance of ${\hat{\theta}}_{n}$, the
deterministic term $p^{-2}\bfone'\Si\bfone$
has no effect, but the $O_p(n^{-1/2})$-term in \eqref{col-rep}
has a nontrivial contribution
to the bias of the resulting column-based
variance estimator, which can \textit{not} be made negligible.
As a result, this alternative approach fails to produce a
consistent estimator for fixed $p$. In general,
caution must be exercised while applying the column-wise
method for small~$p$.

\section{Theoretical results}\label{sec4}
\subsection{Consistency of Gap Bootstrap I estimator}\label{sec4.1}
The Gap Bootstrap I estimator
$\widehat{\var}_{\mathrm{GP}\mbox{-}\mathrm{I}}({\hat{\theta}}_{n})$
of the (asymptotic) variance matrix of ${\hat{\theta}}_{n}$
is consistent under fairly mild conditions, as stated in
Appendix~\ref{pf-t4.1}. Briefly, these conditions
require (i) homogeneity of pairwise
distributions of the centered and scaled estimators
$\{m^{1/2}(\hth_{jn} - \te)\dvtx1\leq j\leq p\}
$,
(ii) some moment and weak dependence conditions on the
$m^{1/2}(\hth_{jn} - \te)
$'s, and (iii) $p\rightarrow\infty$ as $n\rightarrow\infty$.
In particular, the rows of $\bbx$ need \textit{not} be
exchangeable. Condition (iii) is needed to ensure
consistency of the estimator of the covariance term(s)
in (\ref{gp-I}), which is defined in terms of the average of
the $p(p-1)$ pair-wise differences $\{\hth_{jn}-\hth_{kn} \dvtx1\leq
j\neq k\leq p\}$. Thus, for employing the
Gap Bootstrap I method in an application,
$p(p-1)$ should not be too small,

The following result asserts consistency of the Gap Bootstrap I
variance (matrix) estimator.

\begin{theorem}\label{th4.1}
Under conditions \textup{(A.1)} and \textup{(A.2)} given in
the \hyperref[app]{Appendix}, as $n\rightarrow\infty$,
\[
n \bigl[\widehat{\var}_{\mathrm{GB}\mbox{-}\mathrm{I}}({\hat{\theta
}}_{n}) - \var(
\bar{\te}_n) \bigr] \raw0\qquad {\mbox{in probability.}}
\]
\end{theorem}

\subsection{Consistency of Gap Bootstrap II estimator}\label{sec4.2}
Next consider the Gap Bootstrap II estimator of the
(asymptotic) variance matrix of ${\hat{\theta}}_{n}$. Consistency
of $\widehat{\var}_{\mathrm{GB}\mbox{-}\mathrm{II}}({\hat{\theta
}}_{n})$ holds here
under suitable regularity conditions on the estimators
$\{\hth_{jn}\dvtx1\leq j \leq p\}$
and the length of the
``gap'' $q$ for a large class of time
series that allows the rows of the array
\eqref{data-array} to have nonidentical distributions.
See the \hyperref[app]{Appendix} for details of the conditions and their
implications. It is worth noting that unlike Gap Bootstrap I,
here the column dimension $p$ need not go to infinity for
consistency.

\begin{theorem}\label{th4.2}
Under conditions \textup{(C.1)--(C.4)}, given in
the \hyperref[app]{Appendix}, as $n\rightarrow\infty$,
\[
n \bigl[\widehat{\var}_{\mathrm{GB}\mbox{-}\mathrm{II}}({\hat{\theta
}}_{n}) - \var({
\hat{\theta}}_{n}) \bigr] \raw0 \qquad{\mbox{in probability.}}
\]
\end{theorem}

\section{Simulation results}\label{sec5}
To investigate finite sample properties of the proposed methods, we
conducted a moderately large simulation study involving different
univariate and multivariate
time series models.
For the univariate case, we considered three models:
\begin{longlist}[(III)]
\item[(I)] Autoregressive (AR) models of order two
($X_t = \mu+Y_t$ where $Y_t= \al_1 Y_{t-1} + \al_2 Y_{t-2}+ W_t$).

\item[(II)] Moving average (MA) models of order two
($X_t=\mu+Y_t$ where $Y_t
=\be_1 W_{t-1} + \be_2 W_{t-2}+ W_t$).

\item[(III)] A periodic time series model ($X_t = \mu_t + W_t$, $W_t=
\sigma\ep_t$),\vadjust{\goodbreak}
\end{longlist}
where $W_t= \sigma\ep_t$ and $\{\ep_t\}$ are i.i.d. random variables with
zero mean and
unit variance.
The parameter values of the AR models are $\al_1=0.8, \al_2 = 0.1$
with constant mean $\mu=0.1$ and with $\sigma=0.2$.
Similarly, for the MA models, we took the MA-parameters as
$\beta_1= 0.3$, $\be_2= 0.5$,
and set $\sigma= 0.2$ and $\mu= 0.1$. For the third model,
the mean of the $X_t$-variables were taken as a periodic
function of time~$t$:
\[
\mu_t= \mu+\cos2\pi t/p + \sin2\pi t/p
\]
with $\mu= 1.0$ and $p\in\{5,10,20\}$
and with $\sigma= 0.2$.
In all three cases, the $\ep_t$ are generated from
two distributions, namely, (i)
$N(0,1)$-distribution and (ii) a centered Exponential (1) distribution,
to compare the effects of
nonnormality on the performance of the two methods.
Note that the rows of the generated $\bbx$ are
identically distributed for models I and II but not for model III.
We considered
six combinations of $(n,p)$ where $n$ denotes the sample
size and $p$ the number of time slots (or the
periodicity). The parameter of interest $\te$
was the population mean
and the estimator ${\hat{\theta}}_{n}$ was taken to be the sample mean.
Thus, the row-wise estimators $\hth_{jn}$ were the sample
means of the row-variables
and the weights in \eqref{eqn-lin-rep} were $w_{jn} = 1/p$
for all $j=1,\ldots,p$.
In all, there are ($3\times2\times6=$) $36$ possible combinations of
$(n,p)$-pairs, the error distributions, and the three models.
To keep the size of the paper to a reasonable length, we shall only
present 3 combinations of $(n,p)$ in the tables, while
we present side-by-side box-plots for all 6 combinations
of $(n,p)$, arranged by the error distributions.
All results are based on $500$ simulation runs.

\begin{figure}

\includegraphics{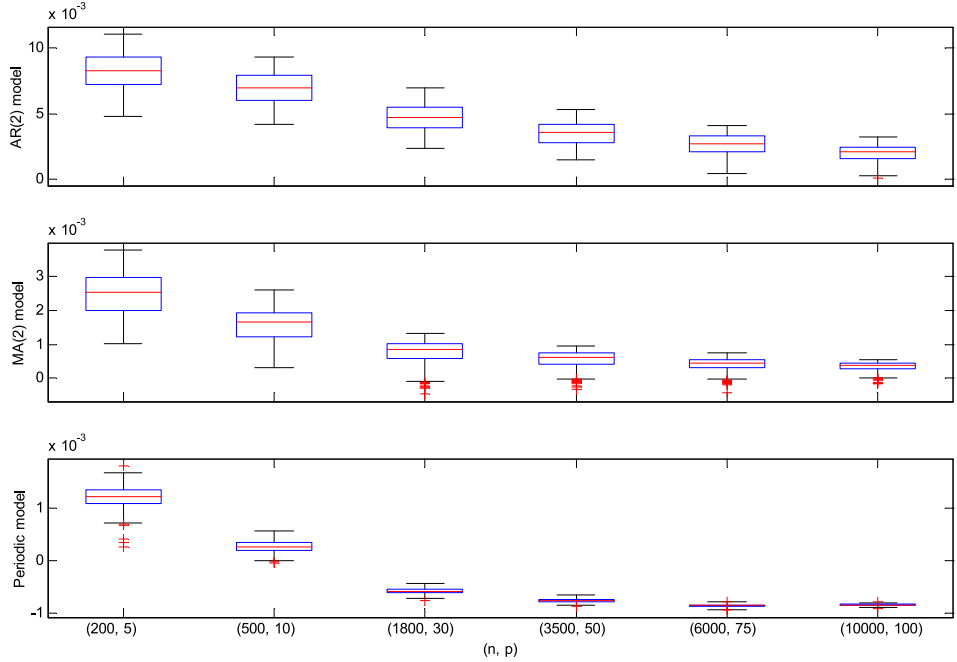}

\caption{Box-plots of the differences between the standard error estimates
based on Gap Bootstrap I and the true standard errors in the one-dimensional
case using $500$ simulation runs.
Here, plots in the first panel are based on
Model I, those in the second and third panels are based on Models II
and III, respectively. The values of $(n,p)$
for each box-plot are given at the bottom of the third
panel. The innovation distribution is centered exponential.}\label{fig2}
\end{figure}

\begin{figure}

\includegraphics{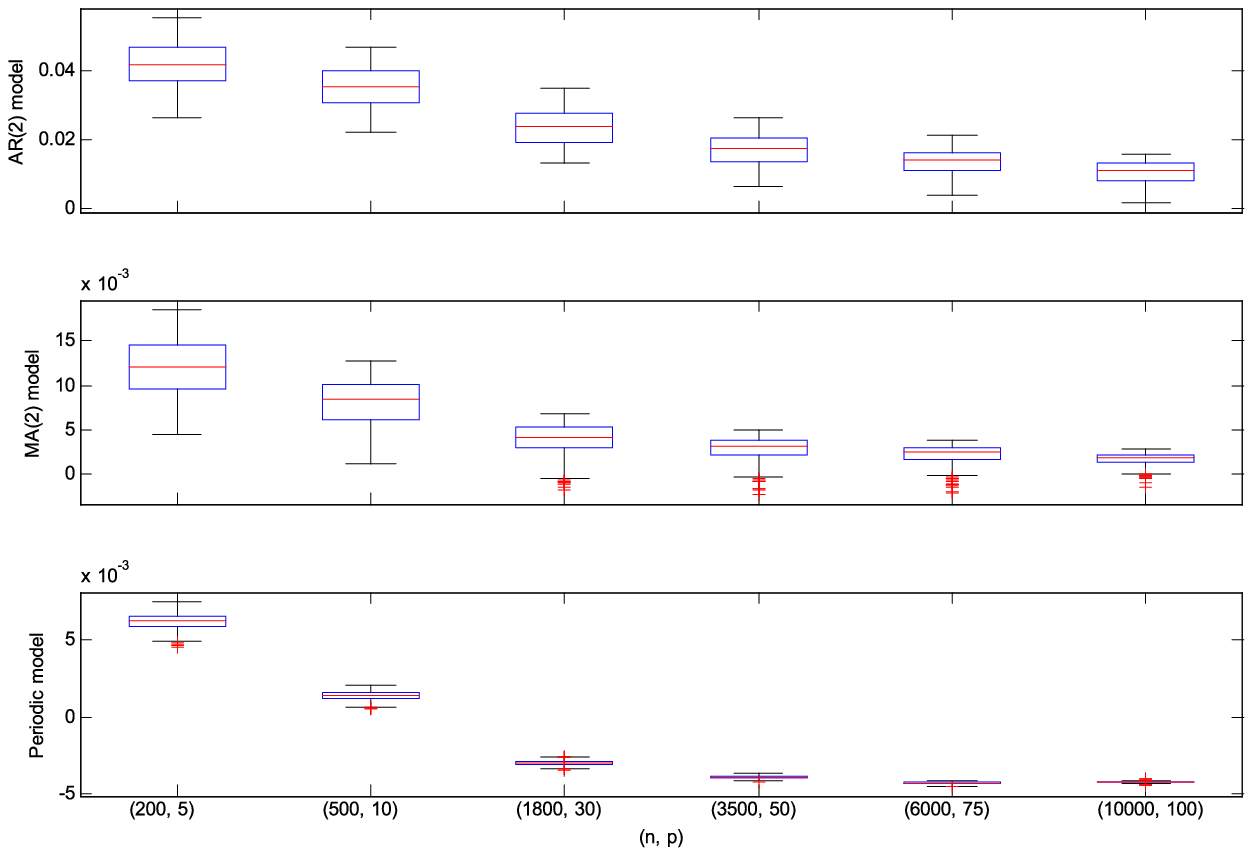}

\caption{Box-plots for the differences of
Gap Bootstrap I estimates and the true standard errors as in
Figure \protect\ref{fig2}, but under normal innovation
distribution.}\label{fig3}
\end{figure}

Figures~\ref{fig2} and~\ref{fig3} give the box-plots of the
differences between the Gap Bootstrap I
standard error estimates and the
true standard errors in the one-dimensional case
under centered exponential and under normal
error distributions, respectively. Here box-plots in the top
panels are based on the $\operatorname{AR}(2)$ model,
the middle panels are based on the $\operatorname{MA}(2)$ model,
while the bottom panels are based on the periodic model.
For each model, the combinations of $(n, p)$ are given by
$(n, p) = (200,5), (500, 10), (1800, 30), (3500, 50),\break
(6000,75), (10\mbox{,}000, 100)$.

%
%

%
%

\begin{figure}

\includegraphics{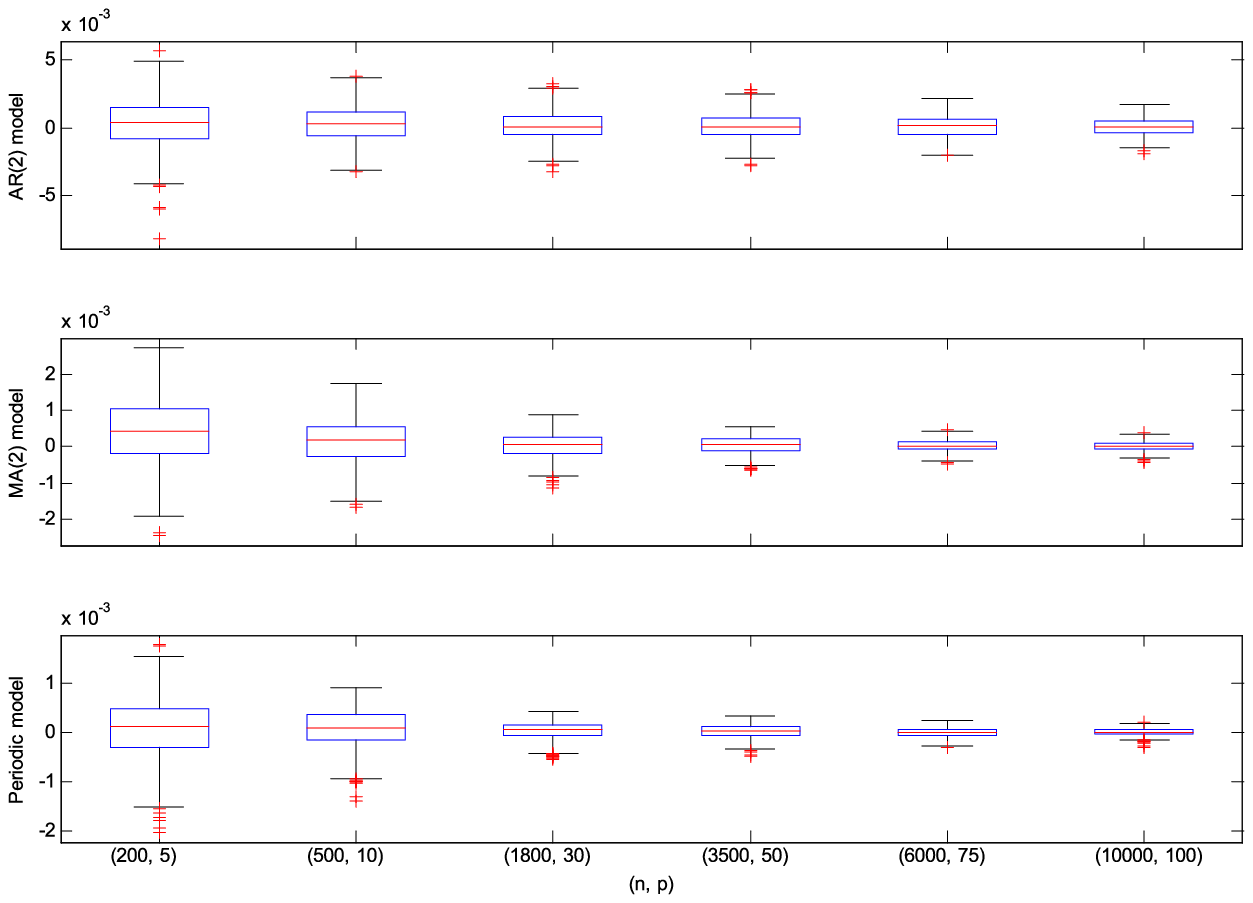}

\caption{Box-plots of the differences
of standard error estimates
based on Gap Bootstrap II
and the true standard errors in the one-dimensional
case, as in Figure \protect\ref{fig2}, under the centered exponential
innovation distribution.}\label{fig4}
\end{figure}

\begin{figure}

\includegraphics{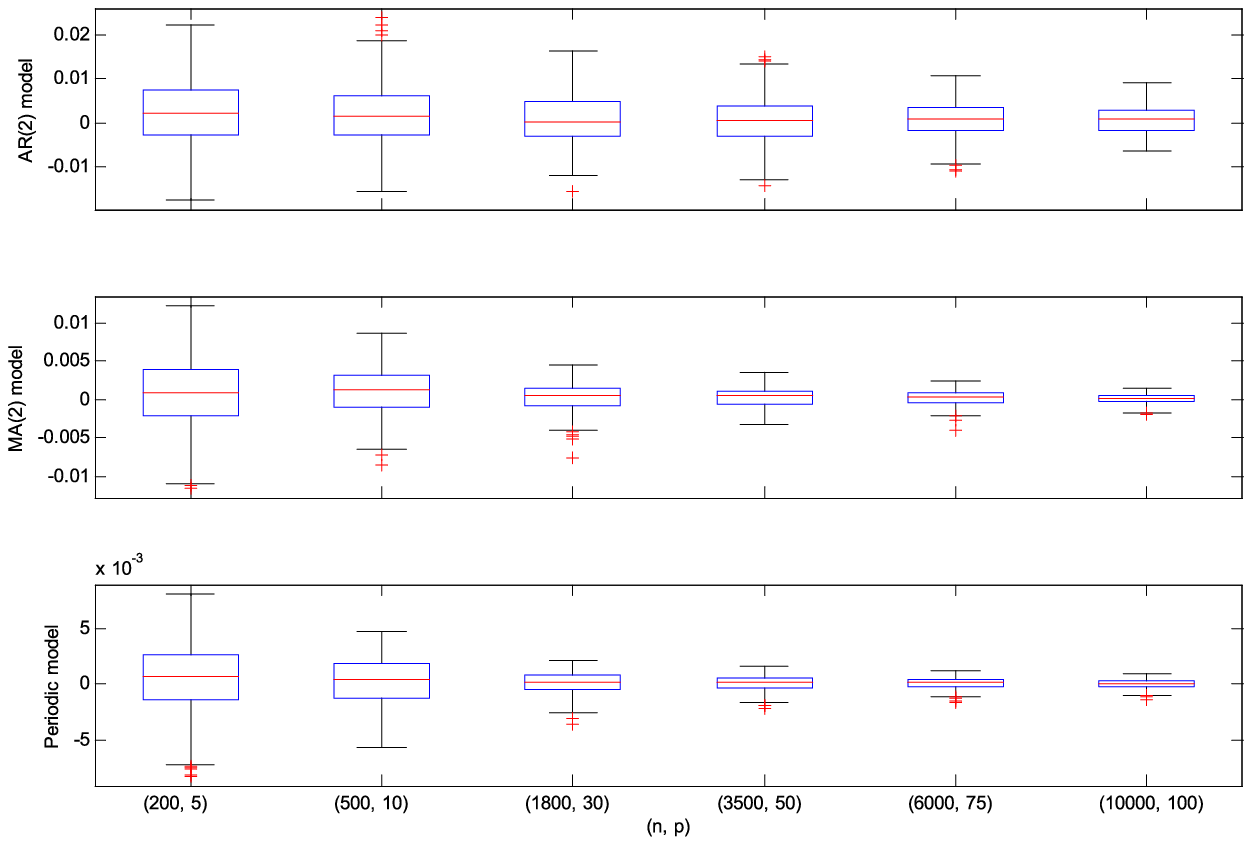}

\caption{Box-plots of the differences
of standard error estimates
based on Gap Bootstrap II
and the true standard errors in the one-dimensional
case, as in Figure \protect\ref{fig2}, under the normal
innovation distribution.}\label{fig5}
\end{figure}

Similarly, Figures~\ref{fig4} and~\ref{fig5} give the
corresponding box-plots for the Gap Bootstrap
II method under centered exponential and under normal
error distributions, respectively.

%
%


%
%


From the Figures~\ref{fig4} and~\ref{fig5}, it is evident that the variability of
the standard error estimates
from the Gap Bootstrap I Method
is higher under Models I and II
than under Model~III for both
error distributions. However, the
bias under Model~III is persistently
higher even for larger values of
the sample size. This can be explained by
noting that for Method I, the assumption of
approximate exchangeability of the rows
is violated under the periodic mean structure
of Model III, leading to a bigger bias.
In comparison, Gap Bootstrap II estimates
tend to center around the target value
(i.e., with differences around zero)
even for the periodic model. Table~\ref{tab1} gives the true values of the
standard errors of ${\hat{\theta}}_{n}$
based on Monte-Carlo simulation and the
corresponding summary measures
for Gap Bootstrap methods I and II
in 18 out of the 36 cases
[we report only the first 3 combinations of $(n,p)$
to save space. A similar pattern was observed
in the other 18 cases].

%
\begin{table}
\caption{Bias and MSEs of Standard Error estimates from
Gap Bootstraps I and II
for univariate data for Models I--III. For each model,
the two sets of 3 rows correspond to
$(n,p)=
(200,5), (500,10), (1800,30)$ under the normal
(denoted by N in the first column) and
the centered Exponential (denoted by E)
error distributions, respectively.
Here B-I${}={}$Bias of Gap Bootstrap I $\times10^2$,
M-I${} = {}$MSE of Gap Bootstrap I $\times10^4$,
B-II${}= {}$Bias of Gap Bootstrap II $\times10^3$,
and M-II${} = {}$MSE of Gap Bootstrap II $\times10^4$.
%
Column 2 gives the target parameter evaluated by
Monte-Carlo simulations and
the last column is the ratio of
columns 4 and 6}\label{tab1}
\begin{tabular*}{\textwidth}{@{\extracolsep{\fill
}}lcd{2.4}d{2.4}d{2.3}d{1.4}d{2.1}@{}}
\hline
\textbf{Model}& \multicolumn{1}{c}{\textbf{True-se}} & \multicolumn
{1}{c}{\textbf{B-I}} & \multicolumn{1}{c}{\textbf{M-I}} &
\multicolumn{1}{c}{\textbf{B-II}} & \multicolumn{1}{c}{\textbf
{M-II}}&\multicolumn{1}{c@{}}{\textbf{Ratio (fix)}}\\
\hline
I.N.1 & 0.013 & -0.831 & 0.708 & -0.376 & 0.029 & 24.4\\
I.N.2 & 0.011 & -0.700 & 0.503 & -0.118 & 0.0202 & 25.2 \\
I.N.3 & 0.008 & -0.481 & 0.241 & -0.256 & 0.0142 & 17.2\\[3pt]
I.E.1 & 0.065 & -4.18 & 17.8 & -1.97 & 0.623 &
28.6\\
I.E.2 & 0.053 & -3.54 & 12.8 & -1.52 & 0.451 & 28.4\\
I.E.3 & 0.038 & -2.41 & 6.04 & -0.844 & 0.348 & 17.4\\[6pt]
II.N.1 & 0.005 & -0.240 & 0.061 & -0.178 & 0.008 &7.6\\
II.N.2 & 0.003 & -0.154 & 0.026 & -0.122 & 0.004 &6.5\\
II.N.3 & 0.002 & -0.081 & 0.007 & -0.087 & 0.001 & 7.0\\[3pt]
II.E.1 & 0.023 & -1.22\mathrm{E} & 1.59 & -1.18 & 0.183 & 8.9\\
II.E.2 & 0.015 & -0.767 & 0.657 & -0.288 & 0.101 & 6.5\\
II.E.3 & 0.008 & -0.398 & 0.184 & -0.092 & 0.025 & 7.4\\[6pt]
III.N.1 & 0.003 & -0.125 & 0.016 & -0.183 & 0.005 &3.2\\
III.N.2 & 0.002 & -0.0263 & 0.0008 & -0.065 & 0.002 & 0.4 \\
III.N.3 & 0.001 & 0.059 & 0.004 & -0.028 & 0.0004 & 10.0\\[3pt]
III.E.1 & 0.014 & -0.619 & 0.386 & -0.549 & 0.094 &4.1\\
III.E.2 & 0.009 & -0.158 & 0.026 & -0.506 & 0.042 & 0.6\\
III.E.3 & 0.005 & 0.292 & 0.086 & -0.216 & 0.010 & 8.6\\
\hline
\end{tabular*}
\end{table}

From the table, we make the following observations:
\begin{longlist}[(iii)]
\item[(i)] The biases of
the Gap Bootstrap I estimators are consistently higher
than those based on Method II under Models I and II
for both normal and nonnormal errors, resulting
in higher overall MSEs for Gap Bootstrap I estimators.
\item[(ii)]
Unlike under Models I and II,
here the biases of the two methods
can have opposite signs.
\item[(iii)]

From the last column of Table~\ref{tab1} (which gives the ratios of the
MSEs of
estimators based on Methods I and II), it follows
that the Gap Bootstrap II works significantly better
than Gap Bootstrap I for Models I and II.
For Model III,
neither method dominates the other in terms of bias
and/or MSE.
MSE comparison shows a curious
behavior of Method I at $(n,p) = (500,10)$
for the periodic model.
\item[(iv)]
The nonnormality of the $\bfX_t$'s
does not seem to have
significant effects on the relative accuracy
of the two methods.
\end{longlist}

Next we consider performance of the two gap Bootstrap methods
for multivariate data. The models we consider are analogs of (I)--(III) above,
with the general structure
\[
\bfY_t = (0.2,0.3,0.4,0.5)'+ \bfZ_t,\qquad
t\geq1,
\]
where $\bfZ_t$ is taken to be the following:
(IV) a multivariate autoregressive (MAR)
process,
(V) a multivariate moving average (MMA)
process, and
(VI) a multivariate periodic process.
For the MAR process,
\[
\bfZ_t = \Psi\bfZ_{t-1}+ \bfe_t,
\]
where
\[
\Psi= \left[
\matrix{ 0.5&0&0&0
\vspace*{2pt}\cr
0.1&0.6&0&0
\vspace*{2pt}\cr
0&0&-0.2&0
\vspace*{2pt}\cr
0&0.1&0&0.4}\right]
\]
and the $\bfe_{t}$ are i.i.d. $d=4$ dimensional normal random vectors with
mean $0$ and covariance matrix $\Si_0$, where we consider two choices of
$\Si_0$:
\begin{longlist}[(ii)]
\item[(i)] $\Si_0$ is the identity matrix of order $4$;
\item[(ii)] $\Si_0$ has $(i,j)$th element given by
$(-\rho)^{|i-j|}$, $1\leq i,j \leq4$,
with $\rho= 0.55$.
\end{longlist}
For the MMA model, we take
\[
\bfZ_t = \Phi_1 \bfe_{t-1}+
\Phi_2 \bfe_{t-2}+\bfe_{t},
\]
where $\bfe_t$ are as above. The matrix of MA coefficients are given by
\[
\Phi_1 = \left[
\matrix{ 1&0&0&0
\vspace*{2pt}\cr
{}*&2&0&0
\vspace*{2pt}\cr
{}*&*&2&0
\vspace*{2pt}\cr
{}*&*&*&2}\right]
\quad{\mbox{and}}\quad \Phi_2 =
\frac{1}{8} \left[ \matrix{ 1&0&0&0
\vspace*{2pt}\cr
{}*&1&0&0
\vspace*{2pt}\cr
{}*&*&1&0
\vspace*{2pt}\cr
{}*&*&*&1}
\right],
\]
where, in both $\Phi_1$ and $\Phi_2$,
the $*$'s are generated by using a random sample from the
UNIFORM $(0,1)$ distribution [i.e., random numbers in $(0,1)$]
and are held fixed throughout the simulation.
We take $\Phi_1$ and $\Phi_2$ as lower triangular matrices to mimic the
structure of the OD model for the real data example that will be considered
in Section~\ref{sec6} below. Finally, the observations $\bfX_t$ under the
periodic model (VI) are
generated by stacking the univariate case with the same $p$,
but
with $\mu$ changed to the the vector
$(0.2,0.3,0.4,0.5)$. 
The component-wise values of $\al_1$ and $\al_2$ are kept the same
and the $\varepsilon_t$'s for the $4$ components are now
given by the $\bfe_t$'s, with the two choices of the
covariance matrix.

\begin{figure}

\includegraphics{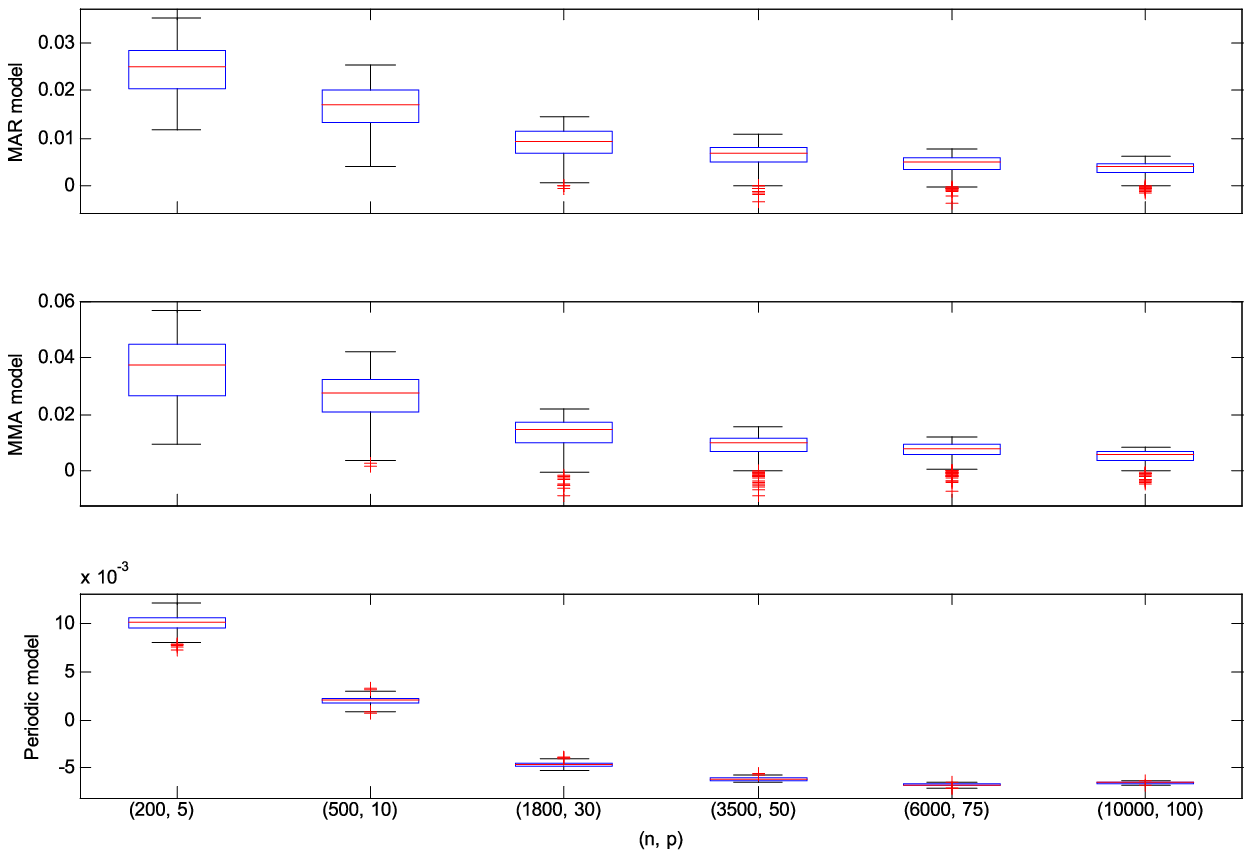}

\caption{Box-plots of the differences
of standard error estimates
based on Gap Bootstrap I
and the true standard errors in the multivariate case,
under the Type II
error distribution.
The number of simulation runs is 500. Also, the
models and the values of $(n,p)$ are depicted
on the panels as in Figure \protect\ref{fig2}.}\label{fig6}
\end{figure}

\begin{figure}

\includegraphics{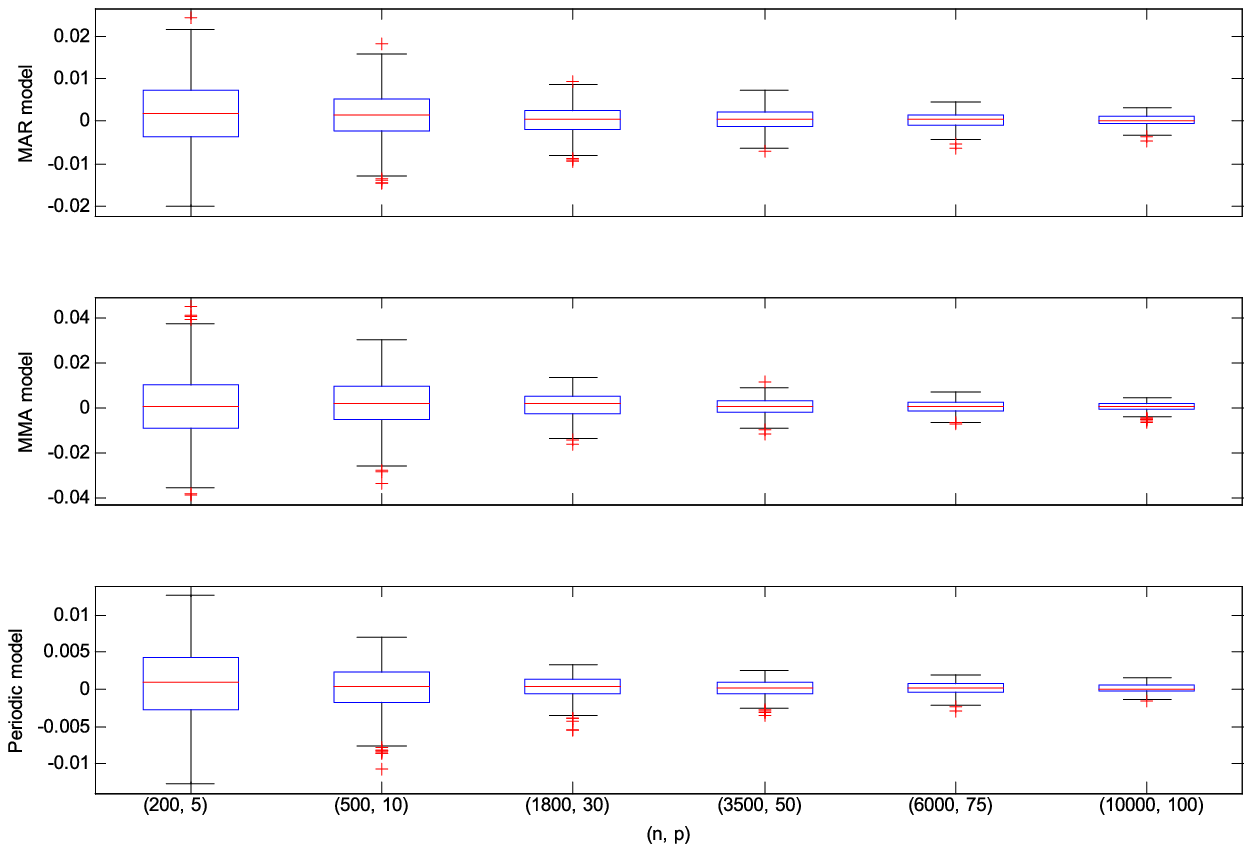}

\caption{Box-plots of the differences
of standard error estimates
based on Gap Bootstrap II
and the true standard errors in the multivariate case,
under the setup of Figure \protect\ref{fig6}.}\label{fig7}
\end{figure}

The parameter of interest is the mean of component-wise means,
that is,
\[
\te=\bar{\mu} = [0.2+0.3+0.4+0.5]/4.
\]
The estimator ${\hat{\theta}}_{n}$ is the mean of the component-wise means
of the \textit{entire} data set
and $\tth^{(i)}$ is given by the mean of the
component-wise means coming from the $i$th row of $n/p$-many
data vectors, for $j=1,\ldots, p$.
Box-plots of the differences between the true standard errors
of ${\hat{\theta}}_{n}$ and their
estimates obtained by
the two Gap Bootstrap methods are reported in Figures
\ref{fig6} and~\ref{fig7}, respectively. We only report the results
for the models with covariance structure (ii) above
(to save space).

%
%


%
%


The number of simulation runs is $500$ as in the univariate case.
From the figures it follows that the relative patterns of the box-plots
mimic those in the case of the univariate case, with Gap Bootstrap
I leading to systematic biases under the periodic mean structure.
For comparison, we have also considered the performance of
more standard methods, namely, the overlapping versions of the
Subsampling (SS) and the Block Bootstrap (BB).

%
%
\begin{figure}

\includegraphics{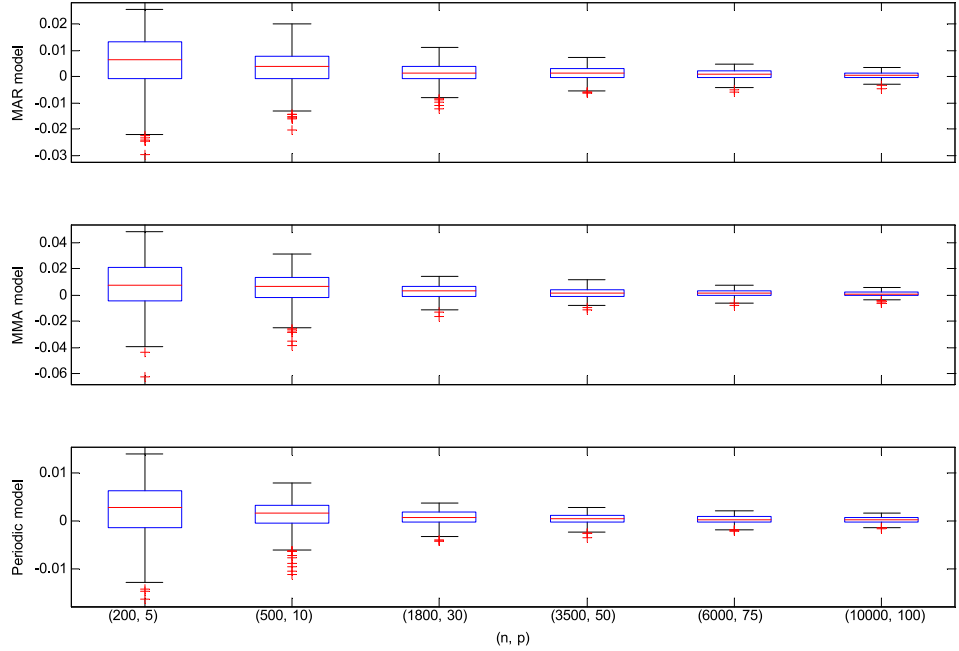}

\caption{Box-plots of the difference
of standard error estimates
based on Subsampling
and the true standard errors in the multivariate case,
under the setup of Figure \protect\ref{fig6}.}\label{fig8}
\end{figure}


%
%
\begin{figure}

\includegraphics{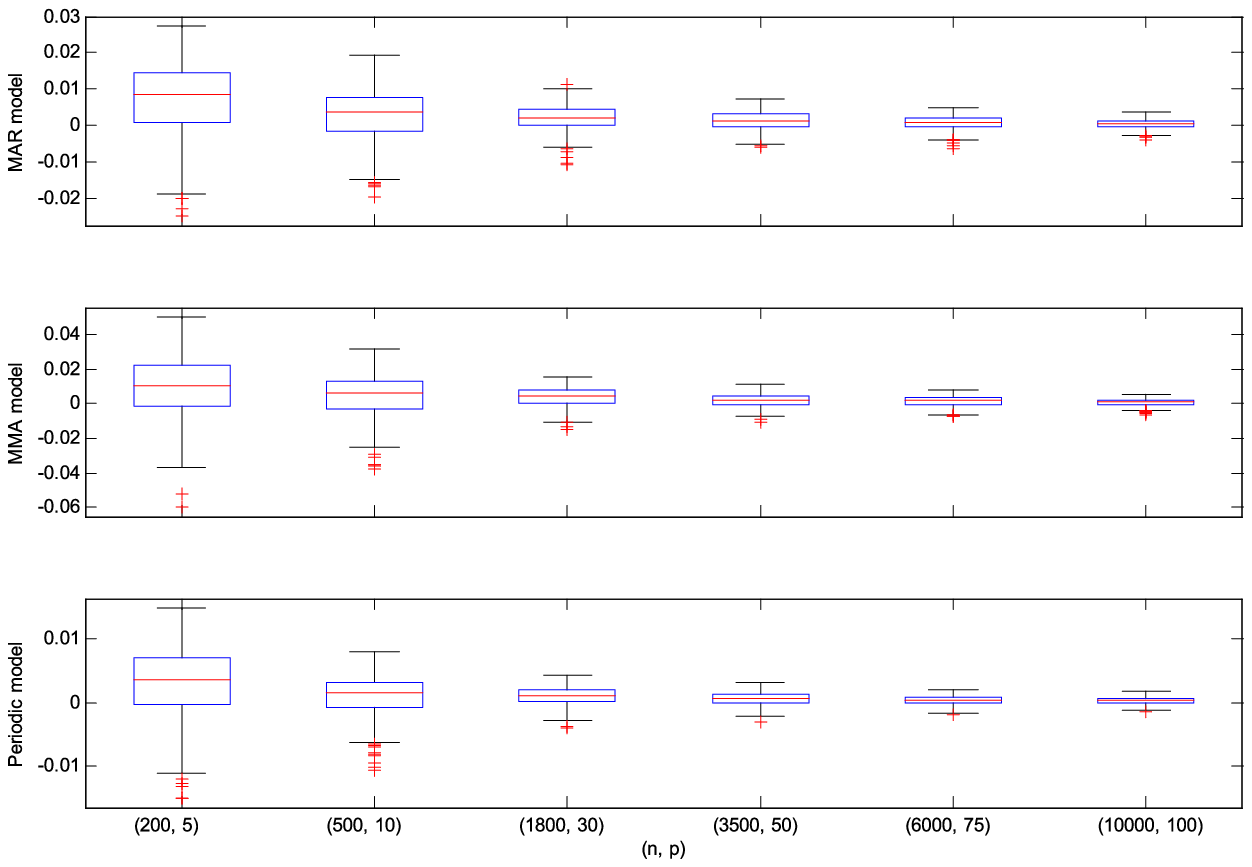}

\caption{Box-plots of the differences
of standard error estimates
based on the block bootstrap
and the true standard errors in the multivariate case,
under the setup of Figure~\protect\ref{fig6}.}\label{fig9}
\end{figure}
%

Figures~\ref{fig8} and~\ref{fig9} give
box-plots of the differences between the true standard errors
of ${\hat{\theta}}_{n}$ and their
estimates obtained by SS and BB methods,\ under Models~(IV)--(VI) with
covariance structure (ii). The choice of the block size was
based on the block length selection rule of Patton, Politis and White
(\citeyear{PatPolWhi09}).
From the figures, it follows that the relative performances of the SS and
the BB methods are qualitatively similar and both methods handily
outperform Gap Bootstrap I.

%
%
\begin{table}
\caption{MSEs of Standard Error estimates from
Gap Bootstraps \textup{I} and \textup{II} and the Subsampling (SS) and
Block Bootstrap (BB)
methods for the multivariate data for Models IV--VI under
covariance matrix of type \textup{(ii)}. The six rows under each model
correspond to
$(n,p)= (200,5), (500,10), (1800,30), (3500,50), (6000,75), (10\mbox{,}000, 100)$.
Further, the entries in the table gives the values of the MSEs multiplied
$10^4$, $10^4$ and $10^5$
for Models~IV--VI, respectively}\label{tab2}
%
%
\begin{tabular*}{\textwidth}{@{\extracolsep{\fill}}lcd{2.3}ccc@{}}
\hline
\textbf{Model}&\textbf{True-se}& \multicolumn{1}{c}{\textbf{GB-I}}&
\textbf{GB-II} &\textbf{SS} & \textbf{BB} \\
\hline
IV.1& 0.044 &6.190 & 0.634& 1.390 & 1.510\\
IV.2&0.030 &2.970 & 0.353& 0.568 & 0.567\\
IV.3&0.017 &0.873 & 0.116& 0.151 & 0.162\\
IV.4&0.012 &0.451 & 0.064& 0.078 & 0.082\\
IV.5&0.009 &0.247 & 0.034& 0.040 & 0.042\\
IV.6&0.007 &0.155 & 0.017& 0.020 & 0.020\\[3pt]
V.1&0.076 & 14.300 & 2.350 & 3.690 & 4.040\\
V.2&0.053 & 7.560 & 1.190 & 1.650 & 1.690\\
V.3&0.028 & 2.060 & 0.300 & 0.374 & 0.427\\
V.4&0.019 & 0.930 & 0.144 & 0.165 & 0.176\\
V.5&0.015 &0.590 & 0.080 & 0.094 & 0.099\\
V.6&0.011 & 0.297 & 0.037 & 0.043 & 0.045\\[3pt]
VI.1&0.022 &10.300 &2.400 &3.150 &3.440\\
VI.2&0.014 &4.250 &0.918 &1.110 &1.100\\
VI.3&0.007 & 2.230 &0.215 &0.257 &0.291\\
VI.4&0.005 & 3.860 &0.111 &0.134 &0.140\\
VI.5&0.004 &4.620 &0.069 &0.073 &0.074\\
VI.6&0.003 &4.350 &0.032 & 0.036 &0.038\\
\hline
\end{tabular*}
\end{table}

These qualitative observations are more precisely quantified in
Table~\ref{tab2} which gives the MSEs of all 4 methods
for models (IV)--(VI) for all six combinations
of $(n,p)$ under covariance structure (ii). It follows from the
table that Gap Bootstrap Method II has the best overall performance
in terms of the MSE. This may appear somewhat counter-intuitive
at first glance, but the gain in efficiency of Gap Bootstrap
II can be explained by noting that it results from judicious choices of
resampling methods for different parts of the target parameter,
as explained in Section~\ref{sec3.3.3} (cf. Remark~\ref{rem3.1}). On the other
hand, in
terms of
computational time, Gap Bootstrap I had the best possible performance, followed
by the SS, Gap Bootstrap~II and the BB methods, respectively. Since the
basic estimator ${\hat{\theta}}_{n}$ is computationally very simple
(being the sample
mean), the computational time may exhibit a very different relative
pattern (e.g., for ${\hat{\theta}}_{n}$ requiring high-dimensional
matrix inversion, the BB method based on the entire data set
may be totally infeasible).

\section{A real data example: The OD estimation problem}\label{sec6}
\setcounter{equation}{0}

\subsection{Data description}\label{sec6.1}

\begin{figure}

\includegraphics{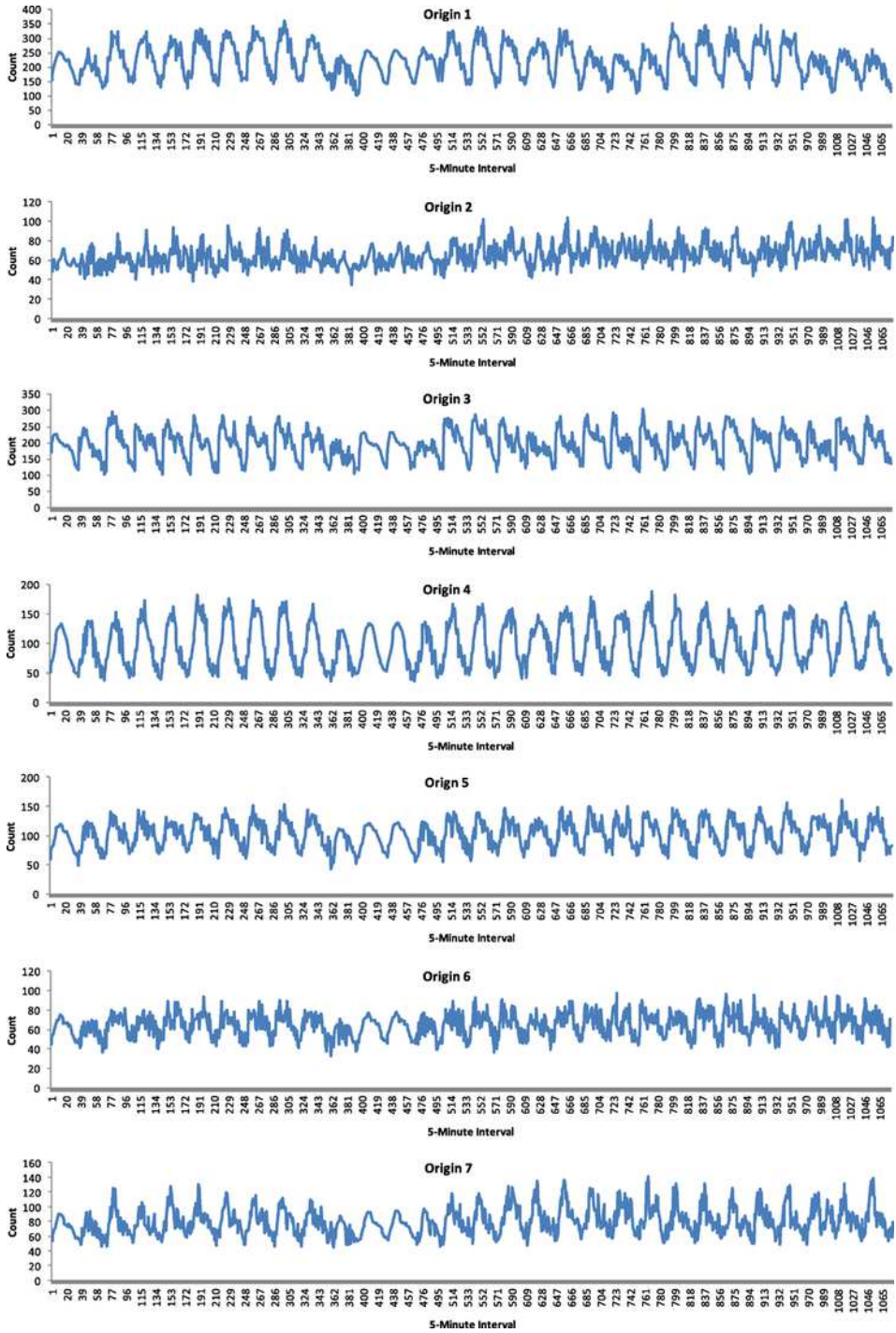}

\caption{Plots of the origin volume counts for
the San Antonio, TX data (including weekend days).}\label{fig10}
\end{figure}

\begin{figure}

\includegraphics{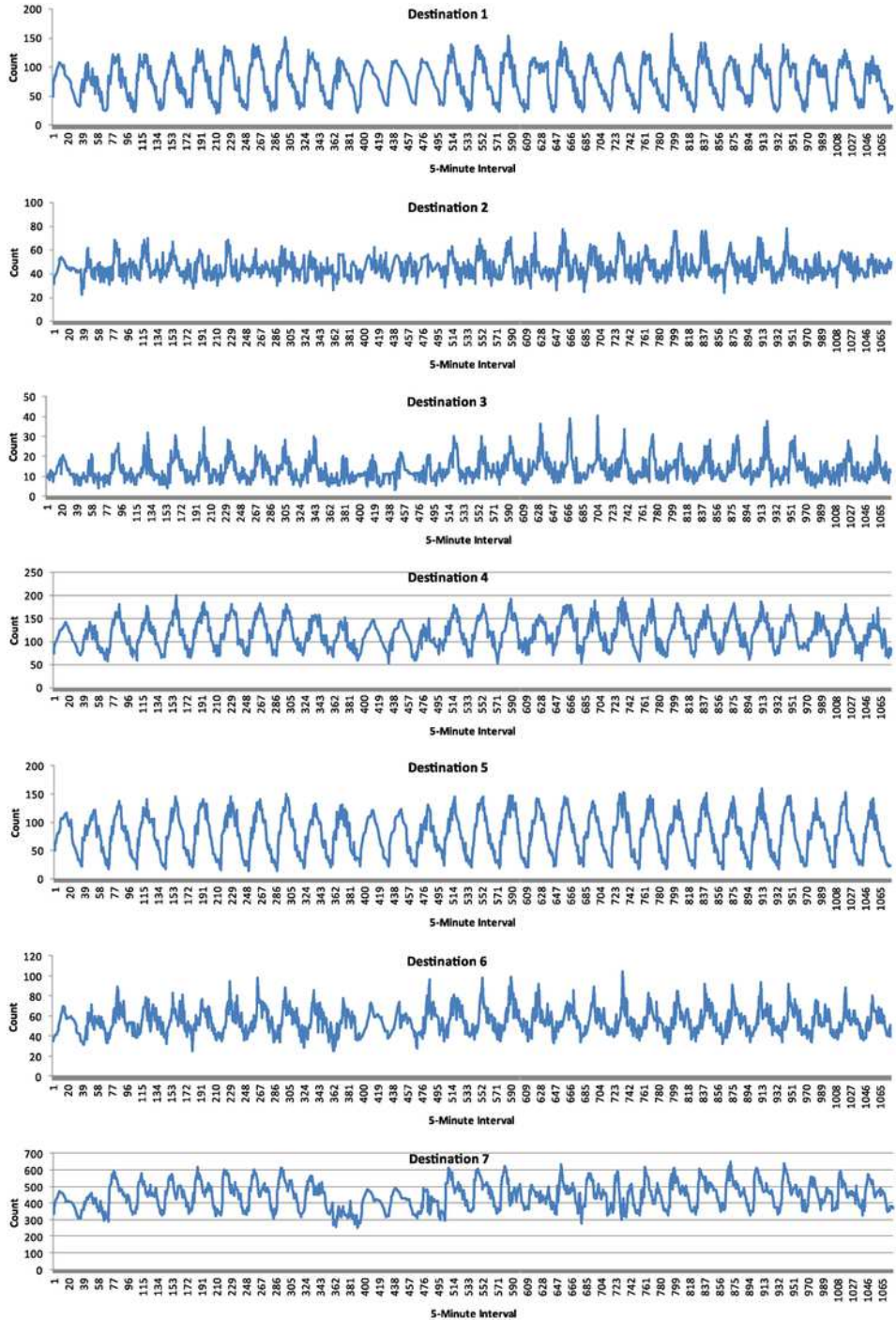}

\caption{Plots of the destination volume counts
for the San Antonio, TX data (including weekend days).}\label{fig11}
\end{figure}

A 4.9 mile section of Interstate 10 (I-10) in San
Antonio, Texas was chosen as the test bed for this study. This section
of freeway is monitored as part of San Antonio's TransGuide Traffic
Management Center, an intelligent transportation systems application
that provides motorists with advanced\vadjust{\goodbreak} information regarding travel times,
congestion, accidents and other traffic conditions. Archived
link volume counts from a series of 14 inductive loop
detector locations (2 main lane locations, 6 on-ramps and 6 off-ramps)
were used in this study (see Figure~\ref{fig1}). The analysis is based
on 575 days of peak AM (6:30 to 9:30) traffic count data (All
weekdays---January 1, 2007 to March 13, 2009). Each day's data were
summarized into 36 volume counts of 5-minute duration. Thus,
there were a total of 20,700 time points,
and each time point giving $14$ origin-destination
traffic data, resulting in \textit{more than a quarter-million
data-values}.
Figures~\ref{fig10} and~\ref{fig11} are plots showing the periodic
behavior of the link volume count data at the 7 origin (O1 to O7)
and 7 destination (D1 to D7) locations, respectively.


\subsection{A synthetic OD model}\label{sec6.2}
As described in Section~\ref{sec2}, the OD trip matrix is required
in many traffic applications such as traffic simulation models,
traffic management, transportation planning and economic
development. However, due to the high cost of direct measurements, the
OD entries are constructed using synthetic OD models
[Cascetta (\citeyear{Cas84}), Bell (\citeyear{Bel91}), Okutani
(\citeyear{Oku87}),
Dixon and Rilett (\citeyear{DixRil00})].
One common approach for estimating
the OD matrix from link volume counts is based on the
least squares regression where the unknown OD matrix
is estimated by minimizing the squared Euclidean distance
between the observed link volumes and the estimated link
volumes.\vadjust{\goodbreak}

Given the link volume counts on all origin and destination ramps, the
OD split
proportion, $p_{ij}$ (assumed homogeneous over the
morning rush-hours), is
the fraction of vehicles that exit the system at destination ramp
$d_{jt}$ given that they enter at origin ramp $o_{it}$ at time point $t$
(cf. Section~\ref{sec2}). Once the split proportions are known, the OD
matrix for
each time period can be identified as a linear combination of the split
proportion matrix and the vector of origin volumes.
It should be noted that because the origin volumes are dynamic, the estimated
OD matrix is also dynamic. However, the split proportions are typically
assumed constant so that the OD matrices by time slice are linear
functions of each other [Gajewski et al. (\citeyear{Gajetal02})]. While
this is a reasonable
assumption for short freeway segments over a time span with homogeneous
traffic patterns
like the ones used in this study, it elicits
the question as to when trips began and ended when used on
larger networks over a longer tie span. It
is also assumed that all vehicles that enter the system from each
origin ramp
during a given time period exit the system during the same time period.
That is,
it is assumed that conservation of vehicles holds, so that the sum of the
trip proportions from each origin ramp equals 1. Caution should be exercised
in situations where a large proportion of trips begin and end during different
time periods [Gajewski et al. (\citeyear{Gajetal02})]. Note also that
some split proportions
such as $p_{21}$ are not feasible because of the structure of the network.
Moreover, all vehicles that enter the freeway from origin ramp 7 go through
destination ramp 7 so that $p_{77}=1$. All of these constraints need to be
incorporated into the estimation process.

Let $d_{jt}$ denote the volume at destination $j$ over the $t$th time interval
(of duration $5$ minutes) and $o_{jt}$ denote the $j$th origin volume
over the same period. Let $p_{ij}$ be the proportion of origin $i$ volume
contributing to the destination $j$ volume (assumed not to change over time).
Then, the synthetic OD model for the link volume counts can be
described as follows:

For each $t$,
%
%
\begin{eqnarray}
\qquad d_{1t}&=& o_{1t}p_{11}+ \ep_{1t},
\nonumber
\\[-1pt]
d_{2t}&=& o_{1t}p_{12}+ o_{2t}p_{22}+
\ep_{2t},
\nonumber
\\[-1pt]
d_{3t}&=& o_{1t}p_{13}+ o_{2t}p_{23}+
o_{3t}p_{33}+ \ep_{3t},
\nonumber
\\[-1pt]
d_{4t}&=& o_{1t}p_{14}+ o_{2t}p_{24}+
o_{3t}p_{34}+ o_{4t}p_{44}+
\ep_{4t},
\\[-1pt]
d_{5t}&=& o_{1t}p_{15}+ o_{2t}p_{25}+
o_{3t}p_{35}+ o_{4t}p_{45}+
o_{5t}p_{55}+ \ep_{5t},
\nonumber\\[-1pt]
d_{6t}&=& o_{1t}p_{16}+ o_{2t}p_{26}+
o_{3t}p_{36}+ o_{4t}p_{46}+
o_{5t}p_{56}+ o_{6t}p_{66}+
\ep_{6t},
\nonumber
\\[-1pt]
d_{7t}&=& o_{1t}p_{17}+ o_{2t}p_{27}+
o_{3t}p_{37}+ o_{4t}p_{47}+
o_{5t}p_{57}+ o_{6t}p_{67}\nonumber\\[-1pt]
&&{}+
o_{7t}p_{77}+ \ep_{7t},\nonumber\vadjust{\goodbreak}
\end{eqnarray}
where $\ep_{jt}$ are (correlated) error variables. Note that the parameters
$p_{ij}$ satisfy the conditions
%
%
\begin{equation}
\sum_{j=i}^7 p_{ij}=1\qquad {
\mbox{for }} i=1,\ldots,7. %
\end{equation}
In particular, $p_{77}=1$. Because of the above linear
restrictions on the $p_{ij}$'s, it is enough to
estimate the parameter vector
$\bfp=(p_{11},p_{12},\ldots,p_{16}; p_{22},\ldots,p_{26};\break
\ldots;  p_{66})'$. We relabel the
components and write $\bfp=(\theta^{[1]},\ldots,\theta^{[21]})'
\equiv{\te}$.
We will estimate these parameters by the least squares method
using the entire data, resulting in the estimator ${\hat{\theta}}_{n}$
and using the daily data over each of the 36 time intervals of length
$5$ minutes, yielding $\hth_{jn}$, $j=1,\ldots,24$. For notational
simplicity, we set $\hth_{0n}={\hat{\theta}}_{n}$.

%
%

%


%

For
$t=1,\ldots,20\mbox{,}700 $, let $D_t=(d_{1t},\ldots,d_{6t}, d_{7t} -\sum
_{i=1}^7 o_{1i})'$
and let $O_t$ be the $7\times21$ matrix given by
\[
O_t= \bigl[ %
O_t^{[1]}
\dvtx\ldots\dvtx O_t^{[6]}
\bigr] ,
\]
where, for $k=1,\ldots,6$,
$O_t^{[k]}$ is a $7\times(7-k)$ matrix with its
last row given by $(-o_{kt},\ldots, -o_{kt})$
and the rest of the elements by
\[
\bigl(O_t^{[k]}\bigr)_{ij} = o_{kt}
\ind(i\geq k)\ind(j=i-k+1),\qquad i=1,\ldots,6, j=1,\ldots, 7-k.
\]
For $j=0,1,\ldots,36$, let
%
%
\begin{equation}
\hth_{jn}
=\biggl[\sum_{t\in T_j}
O_t'O_t \biggr]^{-1}\sum
_{t\in T_j} O_t'D_t,
\end{equation}
where
$T_j=\{j,j+36,\ldots,j+(574\times36)\}$ for $j=1,\ldots,36$ and
where $T_{0}=\{1,\ldots,720\}$. Note that each of
$T_1,\ldots,T_{36}$ has size $575$ (the total number of days)
and corresponds to the counts data
over the respective $5$ minute period, while $T_{0}$ has size $20\mbox{,}700$
and it corresponds to the entire data set. For applying Gap Bootstrap II,
we need a minor extension of the formulas given in Section \ref
{sec3.3}, as the weights
in \eqref{eqn-lin-rep} now vary component-wise.
For $j=0,1,\ldots,36$, define
$
\Ga_{jn} = \sum_{t\in T_j} O_t'O_t$.
Then, the following version of \eqref{eqn-lin-rep} holds
[without the $o_p(1)$ term]:
\[
{\hat{\theta}}_{n}= \sum_{j=1}^{36}
W_{jn}\hth_{jn},
\]
where $W_{jn} = \Ga_{0n}^{-1}\Ga_{jn}$.
This can be proved by noting that
\[
{\hat{\theta}}_{n}= \Ga_{0n}^{-1} \sum
_{t\in T_0} O_t'D_t=
\Ga_{0n}^{-1} \sum_{j=1}^{36}
\sum_{t\in T_j} O_t'D_t
\equiv\sum_{j=1}^{36} W_{jn}
\hth_{jn}.
\]
The Gap Bootstrap II
estimator of the variance of the individual
components ${\hat{\theta}}_{n}^{[1]}, \ldots,{\hat{\theta}}_{n}^{[21]}$
of the estimator ${\hat{\theta}}_{n}$
is now given by
\[
\widehat{\var}\bigl({\hat{\theta}}_{n}^{[a]}\bigr)  = \sum
_{k=1}^{36} \sum
_{l=1}^{36} \hat{\sigma}_{ak}\hat{
\sigma}_{al} {\tilde{\rho}}_a(k,l),\qquad a=1,\ldots,21,
\]
where $\hat{\sigma}_{ak}^2 = \bfw_{ak}'\hSi^{(k)}\bfw_{ak}$,
$\hSi^{(k)}$ is the i.i.d. bootstrap based estimator of the variance matrix
of $\hth_{kn}$,
${\tilde{\rho}}_a(k,j)$ is
the sampling window estimator of the correlation
between the $a$th component of the
$k$th and $j$th row-wise estimators of $\te$ and
$\bfw_{ak}$'s are weights based on $W_{jn}$'s. Indeed, with
$\bfe_1=(1,0,\ldots,0)',\ldots, \bfe_{21}=(0,\ldots,1)'$,
we have $\bfw_{aj}=\bfe_{a}'\Ga_{0n}^{-1}\Ga_{jn}$,
$1\leq j\leq36$. To find ${\tilde{\rho}}_a(k,j)$'s,
we applied the sampling window method
estimator with
$\ell=17$ and the following formula for ${\tilde{\rho}}_a(k,j)$:
\[
{\tilde{\rho}}_a(k,j) = \frac{I^{-1}\sum_{i=1}^I
(\bfw_{ak}'[\tth^{(i)}_{kn} -{\hat{\theta}}_{n}] )
(\bfw_{aj}'[\tth^{(i)}_{jn} -{\hat{\theta}}_{n}] )} {
[I^{-1}\sum_{i=1}^I
(\bfw_{ak}'[\tth^{(i)}_{kn} -{\hat{\theta}}_{n}] )^2
]^{1/2}
[I^{-1}\sum_{i=1}^I
(\bfw_{aj}'[\tth^{(i)}_{jn} -{\hat{\theta}}_{n}] )^2
]^{1/2} },
\]
$j,k=1,\ldots,36$, $a=1,\ldots,21$,
where $\tth^{(i)}_{kn}$'s is the $i$th subsample version
of $\hth_{kn}$ and $I=575-\ell+1=559$.
Following the result on the optimal order of
the block size for estimation of (co)-variances
in the block resampling literature
[cf. Lahiri (\citeyear{Lah03})], here we have set $\ell= c N^{1/3}$
with $N=575$ and $c=2$.

Table~\ref{tab3} gives the estimated standard errors of the
least squares estimators of the~$21$ parameters $\te_1,\ldots,\te_{21}$.

%
\begin{table}
\def\arraystretch{0.9}
\caption{Standard Error estimates from Gap Bootstraps I and
II (denoted by STD-I and~STD-II,~resp.)
for the San Antonio, TX data}\label{tab3}
\begin{tabular*}{\textwidth}{@{\extracolsep{\fill}}lccclccc@{}}
\hline
$\bolds{p_{ij}}$& \textbf{Estimates} & \textbf{STD-I} & \textbf{STD-II}
&
$\bolds{p_{ij}}$& \textbf{Estimates} & \textbf{STD-I} & \multicolumn
{1}{c@{}}{\textbf{STD-II}}\\
\hline
$p_{11}$
& 0.355 & 0.0009 & 0.0019
&
$p_{33}$ & 0.046 & 0.0026 & 0.0041\\
$p_{12}$ & 0.104 &0.0018 &0.0042
& $p_{34}$ & 0.232 & 0.0032 & 0.0132\\
$p_{13}$ & 0.011 & 0.0006 & 0.0015
& $p_{35}$ & 0.106 & 0.0061 & 0.0082\\
$p_{14}$ & 0.064 & 0.0043 & 0.0131
& $p_{36}$ & 0.039 & 0.0025 & 0.0080\\
$p_{15}$ & 0.047 & 0.0024 & 0.0073
& $p_{44}$ & 0.436 & 0.0100 & 0.0155\\
$p_{16}$ & 0.022 & 0.0017 & 0.0042
& $p_{45}$ & 0.240 & 0.0123 & 0.0094\\
$p_{22}$ & 0.385 & 0.0079 & 0.0118
& $p_{46}$ & 0.105 & 0.0057 & 0.0141\\
$p_{23 }$ & 0.083 & 0.0044 & 0.0066
& $p_{55}$ & 0.233 & 0.0080 & 0.0130\\
$p_{24}$ & 0.242 & 0.0053 & 0.0237
& $p_{56}$ & 0.109 & 0.0045 & 0.0168\\
$p_{25}$ & 0.112 & 0.0107 & 0.0144
& $p_{66}$ & 0.537 & 0.0093 & 0.0263\\
$p_{26}$ & 0.064 & 0.0037 & 0.0058
&&&&\\
\hline
\end{tabular*}
\end{table}

From the table, it is evident that the estimates generated by Gap
Bootstrap I
are consistently smaller than those produced by Gap Bootstrap II.
To verify the presence of serial correlation within columns,
we also computed the component-wise sample autocorrelation
functions (ACFs) for each of origin and destination
time series (not shown here). From these, we found that
there is nontrivial correlation in all other
series up to lag $14$ and that the ACFs are of different shapes.
In view of the nonstationarity of the components and the presence
of nontrivial serial correlation, it seems reasonable
to infer that Gap Bootstrap I underestimates the standard error of the
split proportion estimates in the synthetic OD model and, hence,
Gap Bootstrap~II estimates may be used for further analysis and
decision making.

\section{Concluding remarks}\label{sec7}
In this paper we have presented two resampling
methods that are suitable for
carrying out inference on a class of massive data sets that have a
special structural property. While naive applications of the
existing resampling methodology are severely constrained by the
computational issues associated with massive data sets, the proposed methods
exploit the so-called ``gap'' structure of massive data sets to split
them into well-behaved smaller subsets where judicious combinations
of known resampling techniques can be employed to
obtain subset-wise accurate solutions. Some simple analytical
considerations are then used to combine the piece-wise
results to solve the\vadjust{\goodbreak} original problem that is otherwise intractable.
As is evident from the discussions earlier, the versions of the
proposed Gap Bootstrap methods require different sets of
regularity conditions
for their validity. Method I requires that the different subsets (in
our notation, rows) have approximately the same distribution
and that the number of such subsets
be large. In comparison, Method II allows for
nonstationarity among the different subsets and does not
require the number of subsets itself to go to infinity.
However, the price paid for a wider range of
validity for Method II is that it requires some analytical
considerations [cf. \eqref{eqn-lin-rep}] and that it uses
more complex resampling methodology. We show that the analytical
considerations are often simple, specifically
for asymptotically linear estimators, which cover a number of commonly
used classes of estimators. Even in the nonstationary setup, such
as in the regression models associated with the real data example,
finding the weights in~\eqref{eqn-lin-rep} is not very difficult.
In the moderate scale simulation of Section~\ref{sec5},
Method II typically outperformed all the resampling
methods considered here, including, perhaps surprisingly,
the block bootstrap on the entire data set; This can be
explained by noting that unlike the block bootstrap method,
Method II crucially exploits the
gap structure to estimate different parts by using
a suitable resampling method for each part separately.
On the other hand,
Method I gives a
``quick and simple'' alternative for massive data sets
that has a reasonably good performance
whenever the data subsets are relatively homogeneous and the
number of subsets is large.

\begin{appendix}\label{app}
\section*{Appendix: Proofs}
\setcounter{equation}{0}
For clarity of exposition, we first give a relatively
detailed proof of Theorem~\ref{th4.2} in Section~\ref{pf-t4.2} and then
outline a proof of Theorem~\ref{th4.1} in Section~\ref{pf-t4.1}.

\subsection{Proof of consistency of Method II}
\label{pf-t4.2}
\subsubsection{Conditions}
Let $\{\bfY_t\}_{t\in\bbz}$ be a $d$-dimensional time series on a
probability space
$(\Om, \F,P)$ with
strong mixing coefficient
\[
\alpha(n) \equiv\sup\bigl\{\bigl |P(A\cap B) -P(A)P(B) \bigr|\dvtx A\in\F_\infty^a,
B\in\F^\infty_{a+n}, a\in\bbz\bigr\}, \qquad n\geq1,
\]
where $\bbz=\{0,\pm1, \pm2,\ldots\}$ and where
$\F_a^b=\sigma\langle\bfY_t\dvtx t\in[a,b]\cap\bbz\rangle$
for $-\infty\leq a\leq b\leq\infty$.
We suppose that
the observations $\{\bfX_t\dvtx t=1,\ldots,n\}$ are obtained from
the $\bfY_t$-series with systematic deletion of $\bfY_t$-subseries of
length $q$, as described in Section~\ref{sec2.2}, leaving a gap of $q$
in between two columns of $\bbx$, that is, $(\bfX_1,\ldots,\bfX_p)
= (\bfY_1,\ldots, \bfY_{p})$, $(\bfX_{p+1},\ldots,
\bfX_{2p} = (\bfY_{p+q+1},\ldots,\bfY_{2p+q})$, etc.
Thus, for $i=0,\ldots, m-1$
and $j=1,\ldots,p$,
\[
\bfX_{ip+j} = \bfY_{i(p+q)+j}.
\]
Further, suppose that the vectorized process $\{(\bfX_{ip+1},\ldots
,\bfX_{(i+1)p})\dvtx i\geq0\}$
is stationary. Thus, the original process $\{\bfY_t\}$ is\vadjust{\goodbreak}
nonstationary, but it has a
periodic structure over a suitable subset of the index set, as is the
case in the
transportation data example. Note that these assumptions are somewhat
weaker than
the requirements in \eqref{proc-c}. Also, for each $j=1,\ldots,p$,
denote the i.i.d. bootstrap observations generated by Efron (\citeyear{Efr79})'s
bootstrap by
$\{\bfX^*_{ip+j}\dvtx i=0,\ldots,m-1\}$ and the bootstrap version of
$\hth_{jn}$ by
$\te^*_{jn}$. Write $E_*$ and $\var_*$ to denote the conditional
expectation and variance of the bootstrap variables.

To prove the consistency of the Gap bootstrap II
variance estimator, we will make use of the following
conditions:

\begin{enumerate}[(C.4)]
\item[(C.1)]
There exist $C\in(0,\infty)$ and $\delta\in(0,\infty)$ such that
for $j=1,\ldots,p$,
\[
E\psi_j(\bfX_j) = 0,\qquad E\bigl|\psi_j(
\bfX_j)\bigr|^{2+\delta}<C
\]
and $\sum_{n=1}^\infty\al(n)^{{\delta}/{(2+\delta)}}<\infty$.
\item[(C.2)]
$[{\hat{\theta}}_{n}- \sum_{j=1}^p w_{jn}\hth_{jn}] = o (n^{-1/2})$
in $L^2(P)$.
\item[(C.3)] \hspace*{3.1pt}(i)
For $j=1,\ldots,p$,
\[
\hth_{jn} = m^{-1}\sum_{i=0}^{m-1}
\psi_{j} (\bfX_{ip+j}) + o\bigl(m^{-1/2}\bigr)\qquad{
\mbox{in }} L^2(P).
\]
\begin{enumerate}[(ii)]
\item[(ii)] For $j=1,\ldots,p$,
\begin{eqnarray*}
\te^*_{jn} &=& m^{-1}\sum_{i=0}^{m-1}
\psi_{j} \bigl(\bfX^*_{ip+j}\bigr)
+ R_{jn}^*\quad{\mbox{and}}\quad E\bigl[E_* \bigl\{R_{jn}^*\bigr\}^2\bigr] = o
\bigl(m^{-1/2}\bigr),
\\
\tth_{jn}^{(i)} &=&\sum_{a=i}^{i+\ell-1}
\psi_{j}(\bfX_{(a-1)p+j}) + o\bigl(\ell^{-1/2}\bigr)\qquad{
\mbox{in }} L^2(P), i=1,\ldots,I.
\end{eqnarray*}
\end{enumerate}

\item[(C.4)]
$q\rightarrow\infty$ and $p\sum_{j=1}^p w_{jn}^2 = O(1)$ as
$n\rightarrow\infty$.
\end{enumerate}

We now briefly comment on the conditions. Condition (C.1) is a standard
moment and mixing condition used in the literature
for convergence of the series
$\sum_{k=1}^\infty\cov(\psi_j(\bfX_j), \psi_j(\bfX_{kp+j}))$
[cf. Ibragimov and Linnik (\citeyear{IbrLin71})].
Condition (C.2) is a stronger form of \eqref{eqn-lin-rep}. It guarantees
asymptotic equivalence of the variances of ${\hat{\theta}}_{n}$ and
its subsample
(row)-based
approximation $\sum_{j=1}^p w_{jn}\hth_{jn}$. Condition (C.3) in turn
allows us to obtain an explicit expression for the asymptotic variance of
$\hth_{jn}$ and, hence, of ${\hat{\theta}}_{n}$. Note that the linear
representation of $\hth_{jn}$ in (C.3) holds for many
common estimators, including $M$, $L$ and $R$ estimators, where the $L^2(P)$
convergence is replaced by convergence in probability. The $L(P)$ convergence
holds for $M$-estimators under suitable monotonicity conditions
on the score function; for $L$ and $R$-estimators, it
also holds under suitable moment condition on $\bfX_j$'s and
under suitable growth conditions on the weight functions.
Condition (C.3)(ii) requires that a linear representation
similar to that of
the row-wise estimator
$\hth_{jn}$ holds for its i.i.d. bootstrap version $\te_{jn}^*$.
If the bootstrap variables $\bfX^*_{ip+j}$ are defined on
$(\Om, \F,P)$ (which can always be done on a possibly enlarged
probability space), then the iterated expectation $E[E_*\{R_{jn}^*\}^2]$
is the same as the unconditional expectation $E\{R_{jn}^*\}^2$, and
the first part of (C.2)(ii) can be simply stated as
\[
\te^*_{jn} = m^{-1}\sum_{i=0}^{m-1}
\psi_{j} \bigl(\bfX^*_{ip+j}\bigr) + o\bigl(m^{-1/2}
\bigr)\qquad{\mbox{in }} L^2(P).
\]
The second part of (C.2)(ii) is an analog of (C.2)(i) for the
subsample versions of the estimators $\hth_{jn}$'s. The remainder term
here is $o(\ell^{-1/2})$, as the subsampling estimators are now based on
$\ell$ columns of $\bfX_t$-variables as opposed to $m$ columns for
$\hth_{jn}$'s. All the representations in condition (C.3) hold
for suitable classes of $M$, $L$ and $R$ estimators, as described above.

Next consider condition (C.4). It requires that the gap between
the $\bfY_t$ variables in two consecutive columns of $\bbx$
go to infinity, \textit{at an arbitrary rate}. This condition
guarantees that the i.i.d. bootstrap of Efron (\citeyear{Efr79}) yields
consistent
variance estimators for the row-wise estimators $\hth_{jn}$'s,
even in presence of (weak) serial correlation.
The second part of condition (C.4) is equivalent to requiring
$w_{jn} = O(1)$ for each $j=1,\ldots,p$, when $p$ is fixed.
For simplicity, in the following we only prove Theorem~\ref{th4.2}
for the case $p$ is fixed. However, in some applications,
``$p\rightarrow\infty$'' may be a more realistic assumption and, in this
case, Theorem~\ref{th4.2} remains valid provided
the order symbols in (C.3) have the rate $o(m^{-1/2})$
uniformly over $j\in\{1,\ldots,p\}$, in addition to the
other conditions.

\subsubsection{Proofs}\label{appa1.2}
Let $\theta^{\dagger}_n = \sum_{j=1}^p w_{jn}\hth_{jn}$ and $\theta
^{\dagger}_{jn} =
m^{-1}\sum_{i=0}^{m-1}
\psi_j(\bfX_{ip+j})$, $j=1,\ldots,p$. Let $K$ denote a generic
constant in $(0,\infty)$ that does not depend on $n$. Also, unless
otherwise specified,
limits in order symbols are taken by letting $n\rightarrow\infty$.

\begin{pf*}{Proof of Theorem~\ref{th4.2}} First we show that
%
%
\begin{equation}
n \Biggl|\var({\hat{\theta}}_{n}) - \sum_{j=1}^p
\sum_{k=1}^p w_{jn}
w_{kn} \cov(\hth_{jn}, \hth_{kn}) \Biggr| = o(1).
\label{A}
\end{equation}
Let $\De_n = {\hat{\theta}}_{n}-\theta^{\dagger}_n$. Note that by
condition (C.2), $E\De_n^2
= o(1)$.
Hence, by the Cauchy--Schwarz inequality, the left side of \eqref{A} equals
\begin{eqnarray*}
&&n \bigl| E({\hat{\theta}}_{n}-E{\hat{\theta}}_{n})^2
- E\bigl(\theta^{\dagger}_n - E \theta^{\dagger}_n
\bigr)^2 \bigr|
\\
&&\qquad\leq 2n \bigl| E\bigl(\theta^{\dagger}_n - E \theta^{\dagger}_n
\bigr) ( \De_n - E\De_n) \bigr| + n \var(\De_n)
\\
&&\qquad\leq 2n \bigl[\var\bigl(\theta^{\dagger}_n\bigr)
\bigr]^{1/2} \bigl(E\De_n^2\bigr)^{1/2}
+ E\De_n^2
\\
&&\qquad= o(1),
\end{eqnarray*}
provided $\var(\theta^{\dagger}_n) = O(1)$.

To see that $\var(\theta^{\dagger}_n) = O(1)$, note that
%
%
\begin{eqnarray}\label{a}
m\var\bigl(\theta^{\dagger}_{jn}\bigr) &=& m^{-1} \var
\Biggl(\sum_{i=0}^{m-1} \psi_j(
\bfX_{ip+j}) \Biggr)
\nonumber
\\
&=& E\psi_j(X_j)^2 + 2m^{-1}\sum
_{k=1}^{m-1}(m-k) E\psi_j(
\bfX_j) \psi_j(\bfX_{kp+j})
\\
&=& E\psi_j(X_j)^2 + o(1)\nonumber
\end{eqnarray}
as, by conditions (C.1) and (C.4),
\begin{eqnarray*}
&&2 m^{-1}\sum_{k=1}^{m-1} (m-k)
\bigl|E\psi_j(\bfY_j) \psi_j(
\bfY_{k(p+q)+j})\bigr|
\\
&&\qquad\leq K \sum_{k=1}^{m-1} \al\bigl(k[p+q]
\bigr)^{{\delta}/{(2+\delta)}} \bigl(E\bigl|\psi_j(\bfX_j)\bigr|^{2+\delta}
\bigr)^{{2}/{(2+\delta)}}
\\
&&\qquad\leq C^{{2}/{(2+\delta)}} K \sum_{k=p+q}^{\infty}
\al(k)^{{\delta}/{(2+\delta)}} = o(1).
\end{eqnarray*}
By similar arguments, for any $1\leq j,k\leq p$,
%
%
\begin{equation}
m\cov\bigl(\theta^{\dagger}_{jn}, \theta^{\dagger}_{kn}
\bigr) = E\psi_j(\bfX_j)\psi_k(
\bfX_k) + o(1). \label{a2}
\end{equation}

Also, by \eqref{a} and conditions (C.3) and (C.4),
\begin{eqnarray*}
n\var\bigl(\theta^{\dagger}_n\bigr) &=& n \sum
_{j=1}^p w_{jn}^2 \var(
\hth_{jn}) + 2n \sum_{1\leq j<k\leq p}|w_{jn}w_{kn}|
\bigl|\cov(\hth_{jn},\hth_{kn})\bigr|
\\
&=& O \Biggl(\Biggl[\sum_{j=1}^p
|w_{jn}|\Biggr]^{2} n m^{-1} \Biggr) =O(1).
\end{eqnarray*}
Hence, \eqref{A} follows.

To complete the proof of the theorem, by \eqref{A}, it now remains to
show that
%
%
\begin{eqnarray}
m\bigl[\hat{\sigma}_{jn}^2 - \var(\hth_{jn})\bigr] &=&
o_p(1), \label{b1}
\\
\hat{\rho}_{n}(j,k) - \rho_{n}(j,k)&=& o_p(1)
\label{b2}
\end{eqnarray}
for all $1\leq j,k\leq p$,
where $\rho_{n}(j,k)$ is the correlation between $\hth_{jn}$ and
$\hth_{kn}$.
First consider \eqref{b1}. Note that by \eqref{a},\vadjust{\goodbreak} $m \var(\hth_{jn})=
E\psi_j(X_j)^2 + o(1)$
and by condition (C.3)(ii),
\[
m\hat{\sigma}_{jn}^2 = m \var_* \Biggl( m^{-1}
\sum_{i=0}^{m-1} \psi_{j} \bigl(
\bfX^*_{ip+j}\bigr) \Biggr) + o_p(1).
\]
By using a truncation argument and the mixing condition (C.4), it is
easy to show that
\[
m^{-1}\sum_{i=0}^{m-1} \bigl[
\psi_{j} (\bfX_{ip+j})\bigr]^r = E \bigl[
\psi_{j} (\bfX_{ip+j})\bigr]^r +
o_p(1),\qquad r=1,2.
\]
Hence, \eqref{b1} follows. Next, to prove \eqref{b2}, note that
by condition (C.3), \eqref{a} and~\eqref{a2},
\[
\rho_{n}(j,k) = \frac{E\psi_j(\bfX_j)\psi_k(\bfX_k)}{[E\psi_j(\bfX_j)^2]^{1/2}
[E\psi_k(\bfX_k)^2]^{1/2}} + o(1)
\]
for all $j,k$. Also, by conditions (C.3)--(C.4) and standard variance
bound under the moment
and mixing conditions of (C.4), for all $j,k$,
\[
I^{-1}\sum_{i=1}^I
\tth_{jn}^{(i)} \tth_{kn}^{(i)} =
I^{-1}\sum_{i=1}^I
\te_{jn}^{\dagger(i)} \te_{kn}^{\dagger(i)}+
o_p\bigl(\ell^{-1/2}\bigr),
\]
where $\te_{jn}^{\dagger(i)}= \sum_{a=i}^{i+\ell-1} \psi_{j}(\bfX_{(a-1)p+j})$,
$i=1,\ldots,I$. The consistency of the sampling window estimator of
$\rho_n(j,k)$
can now be proved by using conditions (C.2), (C.3) and standard results
[cf. Theorem 3.1, Lahiri (\citeyear{Lah03})]. This completes the proof
of \eqref
{b2} and
hence of Theorem~\ref{th4.2}.

\begin{pf*}{Proofs of \eqref{row-rep} and \eqref{col-rep}}
For notational simplicity, w.l.g., we set $\mu=0$. (Otherwise,
replace $X_t$ by $X_t-\mu$ for all $t$ in the
following steps.) Write $\bX_{jn}$ and $\bX^{(k)}$,
respectively, for the
sample averages of the $j$th row and
$k$th column, $1\leq j\leq p$ and $1\leq k\leq m$.
First consider \eqref{row-rep}. Since $\mu=0$, it follows
that for each $j\in\{1,\ldots,p\}$,
\[
\hth_{jn} = m^{-1}\sum_{i=1}^m
X_{(i-1)p+j}^2 - \bX_{jn}^2 =
m^{-1}\sum_{i=1}^m
X_{(i-1)p+j}^2 +O_p\bigl(n^{-1}\bigr).
\]
Since $n=mp$, using a similar argument, it follows that
${\hat{\theta}}_{n}= n^{-1}\sum_{i=1}^n X_{i}^2 +O_p(n^{-1})
= p^{-1} \sum_{j=1}^p \hth_{jn} +O_p(n^{-1})$. Hence,
\eqref{row-rep} holds.

Next consider \eqref{col-rep}. It is easy to
check that
for all $k=1,\ldots,m$,
$
\hth^{(k)} = p^{-1}\sum_{i=1}^p X_{(k-1)p+i}^2 - [\bX^{(k)}]^2
$ and
$E [\bX^{(k)}]^2 = p^{-2}\bfone'\Si\bfone$.
Hence, with $W_k =[\bX^{(k)}]^2 - E [\bX^{(k)}]^2 $,
\begin{eqnarray*}
{\hat{\theta}}_{n}&=& n^{-1}\sum
_{i=1}^n X_{i}^2
+O_p\bigl(n^{-1}\bigr)
\\
&=& m^{-1} \sum_{k=1}^m
\bigl[ \hth^{(k)} + \bigl\{\bX^{(k)}\bigr\}^2
\bigr] +O_p\bigl(n^{-1}\bigr)
\\
&=& m^{-1} \sum_{k=1}^m
\hth^{(k)} + p^{-2}\bfone'\Si\bfone+
m^{-1} \sum_{k=1}^m
W_k +O_p\bigl(n^{-1}\bigr)
\\
&=& m^{-1} \sum_{k=1}^m
\hth^{(k)} + p^{-2}\bfone'\Si\bfone
+O_p\bigl(n^{-1/2}\bigr),
\end{eqnarray*}
provided condition (C.1) holds with $\psi_j(x) = x^2$ for all $j$.
Further, note that the leading part of the $O_p(n^{-1/2})$-term
is $n^{-1/2}\times m^{-1/2}\sum_{k=1}^m W_k$ and
$m^{-1/2}\times \sum_{k=1}^m W_k$ is asymptotically normal with mean zero
and variance $\sigma_W^2 \equiv\var(W_1)+2\sum_{i=1}^\infty
\cov(W_1, W_{i+1})$. As a result, the $O_p(n^{-1/2})$-term
cannot be of a smaller order (except in the
special case of $\sigma_W^2 = 0$).
\end{pf*}

\subsection{Proof of consistency of Method I}
\label{pf-t4.1}

\subsubsection{Conditions}
We shall continue to use the notation and conventions of Section~\ref{appa1.2}.
In addition to assuming that $\bbx$ satisfies \eqref{proc-c},
we shall make use of the following conditions:

\begin{enumerate}[(A.2)]
\item[(A.1)] \hspace*{2.8pt}(i) Pairwise distributions of $\{m^{1/2}(\hth_{jn} - \te
)\dvtx1\leq j\leq p\}$
are identical.
\begin{enumerate}[(ii)]
\item[(ii)] $\{m^{1/2}(\hth_{jn} - \te)\dvtx1\leq j\leq p\}$ are $m_0$-dependent
with $m_0 = o(p)$.
\end{enumerate}
\item[(A.2)] \hspace*{5.8pt}(i)
$ m\var(\hth_{1n}) \raw\Si$ and $m\cov(\hth_{1n}, \hth_{2n})
\raw\La$ as $n\rightarrow\infty$.
\begin{enumerate}[(iii)]
\item[(ii)] $\{ [m^{1/2}(\hth_{1n} -\te)]^2\dvtx n\geq1\}$ is uniformly
integrable.

\item[(iii)]
$m p^{-1}\sum_{j=1}^p \widehat{\var}(\hth_{jn}) \raw_p \Si$ as
$n\rightarrow\infty$.
\end{enumerate}
\end{enumerate}

Now we briefly comment on the conditions. As indicated earlier,
for the validity of the Gap Bootstrap I method, we do not need
the exchangeability of the rows of $\bbx$; the amount of
homogeneity of the centered and scaled row-wise estimators
$\{m^{1/2}(\hth_{jn} - \te)\dvtx1\leq j\leq p\}$, as specified
by condition (A.1)(i), is all that is needed. (A.1)(i)
also provides the motivation behind the definition of the
variance estimator of the pair-wise differences right above \eqref{gp-I}.
Condition (A.1)(ii) has two implications. First, it quantifies
the approximate independence condition in \eqref{proc-c}. A suitable
strong mixing condition can be used instead, as in the proof of Theorem~\ref{th4.2},
but we do not attempt such generalizations to keep the proof short.
A second implication of (A.1)(ii) is that $p\raw\infty$ as
$n\rightarrow\infty$,
that is, the number of subsample estimators $\hth_{jn}$'s must be large.
In comparison, $m_0$ may or may not go to infinity with $n\rightarrow
\infty$.
Next consider condition (A.2).
Condition (A.2)(i) says that the row-wise estimators are
root-$m$ consistent and that for any pair $j\neq k$,
the covariance between
$m^{1/2}(\hth_{jn} - \te)
$
and
$
m^{1/2}(\hth_{kn} - \te)
$
has a common limit, which is what we are indirectly trying to estimate using
$m\widetilde{\var}(\hth_{jn}- \hth_{kn})$. Condition (A.2)(ii)
is a uniform integrability condition that is implied by
$E| m^{1/2}(\hth_{1n} -\hth_{2n})|^{2+\delta} = O(1)$ [cf. condition (C.1)]
for some $\delta>0$.
Part (iii) of condition (A.2)
says that the i.i.d. bootstrap variance estimator applied to the (average
of the)
row-wise estimators be consistent. A proof of this can be easily constructed
using the arguments given in the proof of Theorem~\ref{th4.2}, by requiring some
standard regularity conditions on the score functions that define the
$\hth_{jn}$'s in Section~\ref{sec3.2}.
We decided to state it as a high level condition to avoid
repetition of similar arguments and to save space.

\subsubsection{\texorpdfstring{Proof of Theorem \protect\ref{sec4.1}}
{Proof of Theorem 4.1}}
In view of condition (A.2)(iii) and \eqref{gp-I}, it is enough to show that
\[
m \bigl[\widetilde{\var}(\hth_{1n} - \hth_{2n}) - E(
\hth_{1n} - \hth_{2n}) (\hth_{1n} -
\hth_{2n})' \bigr] \raw_p 0.
\]
Since this is equivalent to showing component-wise
consistency, without loss of generality, we may suppose that
the $\hth_{jn}$'s are one-dimensional.\break Define
$V_{jk} = m(\hth_{jn} -\hth_{kn})^2\ind(|m^{1/2}(\hth_{jn} -\hth_{kn})| >a_n)$,
$W_{jk} = m(\hth_{jn} -\break\hth_{kn})^2\ind(|m^{1/2}(\hth_{jn} -\hth
_{kn})| \leq a_n)$,
for some $a_n\in(0,\infty)$ to be specified later. It is now enough
to show that
\begin{eqnarray*}
Q_{1n}&\equiv& p^{-2}\sum_{1\leq j\neq k\leq p}
|V_{jk} - EV_{jk}| \raw_p0,
\nonumber
\\
Q_{2n}&\equiv& p^{-2} \biggl|\sum_{1\leq j\neq k\leq p}
[W_{jk} - EW_{jk}] \biggr| \raw_p0. \label{vdiff-c}
\end{eqnarray*}
By condition (A.2)(ii),
$\{[m^{1/2}(\hth_{1n} -\hth_{2n})]^2\dvtx n\geq1\}$ is also uniformly integrable
and, hence,
\[
EQ_{1n} \leq2 E|m^{1/2}(\hth_{1n} -
\hth_{2n})^2 \ind\bigl(\bigl|m^{1/2}(
\hth_{1n} -\hth_{2n})\bigr| >a_n \bigr) = o(1)
\]
whenever $a_n\raw\infty$ as $n\rightarrow\infty$.
Next consider $Q_{2n}$. Define the sets $J_1 = \{(j,k)\dvtx1\leq j\neq
k\leq p\}$,
$A_{j,k}= \{(j_1,k_1)\in J_1\dvtx\min\{|j-j_1|, |k-k_1|\} \leq m_0\}$
and $B_{j,k}
= J_1\setminus A_{j,k}$, $(j,k)\in J_1$. Then, for any $(j,k)\in J_1$,
by the $m_0$-dependence condition,
\[
\cov(W_{jk}, W_{a,b}) = 0\qquad{\mbox{for all }} (a,b) \in
B_{j,k}.
\]
Further, note that $|A_{j,k}|\equiv$ the size of $A_{j,k}$ is at most $2m_0p$
for all $(j,k)\in J_1$. Hence, it follows that
\begin{eqnarray*}
EQ_{2n}^2 &\leq& p^{-4} \biggl[\sum
_{(j,k)\in J_1} \var(W_{jk}) +\sum
_{(j,k)\in J_1} \sum_{(a,b)\neq(j,k)}
\cov(W_{jk}, W_{ab}) \biggr]
\\
&\leq& p^{-4} \biggl[ p^2 EW_{12}^2
+ \sum_{(j,k)\in J_1} \sum_{(a,b)\in A_{j,k}}
\bigl|\cov(W_{jk}, W_{ab})\bigr| \biggr]
\\
&\leq& p^{-4} \bigl[ p^2 a_n^2
E|W_{12}| + p^2 \cdot2m_0p\cdot
a_n^2 E|W_{12}| \bigr]
\\
&=& O\bigl(p^{-1}m_0a_n^2\bigr)
\end{eqnarray*}
as $ E|W_{12}| \leq mE(\hth_{1n}-\hth_{2n})^2 = O(1)$. Now choosing
$a_n = [p/m_0]^{1/3}$
(say), we get $Q_{kn}\raw_p0$ for $k=1,2$, proving \eqref{vdiff-c}.
This completes the proof of
Theorem~\ref{th4.1}.
\end{pf*}
\end{appendix}


%


%


\section*{Acknowledgments}
The authors thank the referees, the Associate
Editor and the
Editor for a number of constructive suggestions that significantly
improved an earlier draft of the paper.

%

\printaddresses


\end{document}